%% file: main.tex
\documentclass[twocolumn,twocolappendix,nofootinbib]{openjournal}
\usepackage{mathptmx}
\usepackage[T1]{fontenc}
\DeclareRobustCommand{\VAN}[3]{#2}
\let\VANthebibliography\thebibliography
\def\thebibliography{\DeclareRobustCommand{\VAN}[3]{##3}\VANthebibliography}
\usepackage{graphicx}	
\usepackage{amsmath}	
\usepackage{amssymb}	
\usepackage{rotating}

\usepackage{aecompl}
\usepackage{array}
\usepackage{times}
\usepackage{url}
\usepackage{xcolor}
\usepackage{soul}
\usepackage{orcidlink}

\usepackage{hyperref}
\hypersetup{colorlinks=true,citecolor=blue,linkcolor=blue,filecolor=blue,urlcolor=blue}
\usepackage{orcidlink}

\newcommand{\Msun}{{\rm M}_{\odot}}
\newcommand{\deriv}[2]{\frac{{\rm d} #1}{{\rm d} #2}}

\shorttitle{Splashback feature depth and width}
\shortauthors{Yu et al.}

\begin{document}
\journalinfo{The Open Journal of Astrophysics}

\title[Splashback feature depth and width]{Measuring the splashback feature: Dependence on halo properties and
history}
\author
    {{Qiaorong S. Yu$^{1,2}$ \orcidlink{0009-0008-3761-3931},
    Stephanie O’Neil$^{3,4,5}$\orcidlink{0000-0002-7968-2088}, 
    Xuejian Shen$^5$\orcidlink{0000-0002-6196-823X},
    Mark Vogelsberger$^5$}\orcidlink{0000-0001-8593-7692},
    Sownak Bose$^6$\orcidlink{0000-0002-0974-5266},
    Boryana Hadzhiyska$^{7,8,9,10}$,
    Lars Hernquist$^{11}$,
    Rahul Kannan$^{12}$\orcidlink{0000-0001-6092-2187},
    Monica Wu$^{13}$, and
    Ziang Wu$^{13,14}$}
\thanks{E-mail: shannon.qyu@gmail.com}
\affiliation{
    $^1$Mathematical Institute, University of Oxford, Oxford, OX2 6GG, UK\\
    $^2$Department of Physics, University of Oxford, Oxford, OX1 3PU, UK\\
    $^3$Department of Physics \& Astronomy, University of Pennsylvania, Philadelphia, PA 19104, USA\\
    $^4$Department of Physics, Princeton University, Princeton, NJ 08544, USA\\
    $^5$Department of Physics and Kavli Institute for Astrophysics and Space Research,
           Massachusetts Institute of Technology,
           Cambridge, MA 02139, USA\\
    $^6$Institute for Computational Cosmology, Department of Physics, Durham University, South Road, Durham, DH1 3LE, UK\\
    $^7$Physics Division, Lawrence Berkeley National Laboratory, Berkeley, CA 94720, USA\\
    $^8$Berkeley Center for Cosmological Physics, Department of Physics, University of California, Berkeley, CA 94720, USA\\
    $^9$Institute of Astronomy, Madingley Road, Cambridge, CB3 0HA, UK\\
    $^{10}$Kavli Institute for Cosmology, Madingley Road, Cambridge, CB3 0HA, UK\\
    $^{11}$Center for Astrophysics | Harvard \& Smithsonian, 60 Garden Street, Cambridge, MA 02138, USA\\
    $^{12}$Department of Physics and Astronomy, York University, 4700 Keele Street, Toronto, ON M3J 1P3, Canada\\
    $^{13}$Courant Institute of Mathematical Sciences, New York University, 251 Mercer Street, New York, NY 10012, USA \\
    $^{14}$Tandon School of Engineering, New York University, New York, NY 11201, USA
    }




\label{firstpage}

\begin{abstract}

In this study, we define the novel splashback depth $\mathcal{D}$ and width $\mathcal{W}$ to examine how the splashback features of dark matter haloes are affected by the physical properties of haloes themselves.
We use the largest simulation run in the hydrodynamic MillenniumTNG project. 
By stacking haloes in bins of halo mass, redshift, mass-dependent properties such as peak height and concentration, and halo formation history, we measure the shape of the logarithmic slope of the density profile of dark matter haloes.
Our results show that the splashback depth has a strong dependence on the halo mass which follows a power law $\mathcal{D}\propto\left(\log_{10}M\right)^{2.8}$.
Properties with strong correlation with halo mass demonstrate similar dependence.
The splashback width has the strongest dependence on halo peak height and follows a power law $\mathcal{W}\propto\nu^{-0.87}$.
We provide the fitting functions of the splashback depth and width in terms of halo mass, redshift, peak height, concentrations and halo formation time.
The depth and width are therefore considered to be a long-term memory tracker of haloes since they depend more on accumulative physical properties, e.g., halo mass, peak height and halo formation time.
They are shaped primarily by the halo’s assembly history, which exerts a stronger influence on the inner density profile than short-term dynamical processes.
In contrast, the splashback features have little dependence on the short-term factors such as halo mass accretion rate and most recent major merger time.
The splashback depth and width can therefore be used to complement information gained from quantities like the point of steepest slope or truncation radius to characterise the halo's history and inner structure.

\end{abstract}


\section{Introduction}\label{sec:intro}
    \input{introduction}

\section{Methods}\label{sec:methods}
    \input{methods}

 \section{Results}\label{sec:results}
    \input{results}

\section{Conclusions}\label{sec:conclusions}
    \input{conclusions}

\section{Acknowledgements}

We thank Bhuvnesh Jain, Susmita Adhikari, Robyn Sanderson, and Aritra Kundu for helpful discussions.
SO is supported by the National Science Foundation under Grant No. AST-2307787.
SB is supported by the UK Research and Innovation (UKRI) Future Leaders Fellowship [grant number MR/V023381/1].
The computations are performed on the Engaging cluster supported by the Massachusetts Institute of Technology, the FARSC cluster supported by Harvard University and the MPCDF cluster supported by the Max Planck Institute.

We made use of the following software for the analysis:
\begin{itemize}
	\item {\textsc{Python}}: \citet{vanRossum1995}
	\item {\textsc{Matplotlib}}: \citet{Hunter2007}
    \item {\textsc{Colossus}}: \citet{Diemer2018}
	\item {\textsc{SciPy}}: \citet{2020SciPy-NMeth}
	\item {\textsc{NumPy}}: \citet{harris2020array}
    \item {\textsc{Seaborn}}: \citet{Waskom2021}
    \item {\textsc{Astropy}}: \citet{astropy:2022}
\end{itemize}

\section{Data Availability}
The data is based on the MilleniumTNG simulations that are expected to be made publicly available at \url{https://www.mtng-project.org/} \citep{hernandez2023millenniumtng}.
Reduced data are available upon request.

\bibliographystyle{mn2e}
\bibliography{bibliography}

\appendix
\input{appendix_fitting}

\input{appendix_dmo}

\label{lastpage}

\end{document}

%% file: introduction.tex
In the current framework of cosmic structure formation, small density perturbations in the early universe grow under the influence of gravity and form complex structures, including dark matter haloes, filaments and voids \citep{benson2010galaxy}. Among these structures, dark matter haloes serve as the gravitational anchors for galaxies and larger cosmic assemblies \citep{Wechsler2018}. Understanding the boundaries and internal dynamics of these haloes is essential for advancing our knowledge of cosmology and galaxy formation.

Our knowledge of dark matter haloes is inferred from observable properties of galaxies and the surrounding baryonic matter, such as gas in the CGM and IGM.
For this purpose, a number of relations have been developed to probe properties that are not directly observable \citep[e.g.][and references therein]{Wechsler2018}.
These include, for example, the total halo mass as a function of stellar mass \citep[e.g.][]{girelli2020stellar}, supermassive black hole mass from bulge velocity dispersion or luminosity \citep[e.g.][]{Kormendy1995,Gebhardt2000,Gultekin2009}, or total galaxy mass from stellar velocity dispersions or rotational velocities \citep[e.g.][]{McGaugh2000,Zahid2016,Barat2019}.

The evolutionary history of galaxies is also difficult to probe.
Galaxies evolve over billions of years, yet we observe them only at one time.
This presents a problem while studying galaxies since their present-day properties can depend in a non-linear way on properties like the timing of mergers, formation time or accretion rate of haloes.
Simulations have therefore been instrumental in studying galaxies as they evolve \citep[see][for review]{Vogelsberger2020}.

However, halo properties like size, total mass, or accretion rate can be difficult to define even in simulations.
Haloes are embedded within a cosmic web and therefore lack a clearly defined edge.
Traditionally, the size of a halo is defined by identifying a region of space where the enclosed density is a fixed multiple of the average density of the universe \citep{Navarro1996}.
This leads to common radial definitions like $R_{\rm200,mean}$ or $R_{\rm200,crit}$ where the enclosed density is 200 times the mean or critical density of the universe.
However, these definitions suffer from effects like pseudo-evolution, the apparent change in size due to the changing background density rather than change intrinsic to the halo \citep{Diemer2013}.
As the background density decreases with the expansion of the universe, the halo radius must correspondingly increase such that the enclosed density decreases to match the background density.
This further complicates measurements of properties such as accretion rate, which relies on measuring the change in mass over time.

The splashback radius $R_{\rm sp}$, marking the outermost edge of a halo where infalling matter reaches its first apocentre, offers a physically motivated boundary distinct from traditional spherical overdensity definitions \citep{adhikari2014splashback, diemer2014dependence}.
This has been identified in a variety of ways. 
\citet{Diemer2017a} and \citet{Diemer2020a} traced the trajectories of individual dark matter particles as they fall into haloes, selecting a radius that encloses the first apocentre of a percentage of their orbits. The smoothed average of the apocentre radii of individual particles is then defined as the splashback radius. 
\citet{diemer2014dependence}, \citet{more2015splashback}, \citet{more2016detection}, and \citet{O'Neil2021,O'Neil2022,O'Neil2024} instead used the steepest point in a spherically averaged density profile as the feature identifying the splashback radius. \citet{mansfield2017splashback} sampled the density field around individual haloes to identify a ‘shell’ around a halo based on the steepest point in the density profile along various lines of sight. This creates a non-spherical boundary instead of defining a single radius. 

In idealised spherical collapse models, the density caustic arises from the pile-up of infalling shells near their first apocentre, where the apocentre of each particle’s first orbit occurs at roughly the same radius \citep{huss1999universal,adhikari2014splashback}. This manifests itself as a minimum of the density slope profile \citep{diemer2014dependence, adhikari2014splashback}, although there is not a one-to-one match in practice.
This has led to many studies using the point of steepest slope, $R_{\rm st}$, as a more easily measured proxy for measuring $R_{\rm sp}$.
However, the relationship between $R_{\rm sp}$ and $R_{\rm st}$ is not exact, and there remains some debate over the use of $R_{\rm st}$ as a measurement for $R_{\rm sp}$ \citep{Diemer2020a}.
Additionally, the steepest slope of a density profile does not necessarily indicate the presence of the splashback radius; any halo will have a steepest point in its density profile. 
The splashback feature is associated with the orbit of matter in a halo and where the logarithmic slope decreases significantly, typically below a value of $-3$ as described in an NFW profile \citep{Navarro_1997}.

The feature has also been detected using galaxies as a tracer of dark matter haloes in clusters \citep[e.g.][]{more2016detection, Baxter2017, Chang2018, Shin2019, Murata2020}.
Because it is not possible to track the trajectories of galaxies in observations, these studies use $R_{\rm st}$ of the galaxy number density as a proxy for $R_{\rm sp}$.
In addition, the galaxy profile does not perfectly align with the dark matter density profile, especially depending on the population of galaxies used \citep{O'Neil2022}.
Observations also view clusters in projection.
Although effects from projecting the profiles can be mitigated with proper fitting, there is still a decrease in the value of $R_{\rm st}$ obtained from profiles in projection compared to 3D profiles \citep{Sun2025}.

The location of $R_{\rm st}$ encapsulates information about, for example, halo accretion histories and dynamical interactions \citep{adhikari2014splashback,diemer2014dependence}. Prior studies have primarily explored the relation between the splashback radius and the halo mass, halo accretion rate, as well as baryonic effects.
However, the ``splashback feature'', which we use to refer to the region of the density profile that rapidly steepens and then transitions to the mean density of the universe, may also contain information about the halo.
Although properties like the shallowness of the splashback feature have been primarily cited as sources of error in measuring $R_{\rm st}$, the sharpness of the feature is affected by factors like the formation history of the halo.

The depth and width of the splashback feature, describing the sharpness and extent of the density gradient at the halo edge, may provide a comprehensive characterisation of halo boundaries.
However, they remain largely unexamined in the context of their dependence on halo properties, growth and environmental influences. The evolution of these features with redshift could also illuminate changes in halo dynamics and the surrounding cosmic web.
\citet{Zhang2023}, for example, noted the splashback feature can impact cluster mass estimates using weak lensing.

In this study, we analyse the depth and width of the splashback feature across a wide range of halo masses and redshifts using the MillenniumTNG simulations. By examining these features in stacked density profiles, we aim to uncover trends and dependencies previously unexplored.

The rest of this paper is structured as follows. In Section \ref{sec:methods}, we describe the simulations, halo selection, the definition of the splashback depth and width, and methods for identifying splashback features. In Section \ref{sec:results_mass}, we examine the splashback features as a function of halo mass and redshift. We look further into the splashback feature's dependence on halo accretion rate and halo merger history in Sections \ref{sec:results_accret} and \ref{sec:results_assembly} including halo peak height (\ref{sec:results_ph}), concentration (\ref{sec:results_conc}) as well as formation time (\ref{sec:results_formz}). We summarise our conclusions in Section \ref{sec:conclusions}.

%% file: methods.tex
\subsection{Simulations}
    \input{methods/methods_simulations}
    
\subsection{Halo properties}
We investigate the dependence of splashback features on the physical properties of dark matter haloes. This includes the halo mass, mass accretion rate, and mass-dependent properties such as the peak height and concentration, as well as halo formation history. In this section, we clarify the definitions of the physical quantities that we use in the study.  
\input{methods/methods_halos}

\subsection{Halo samples}
\label{sec:halo_finder}
    \input{methods/methods_samples}

\subsection{Fitting}\label{sec:fitting}
To identify the splashback features of dark matter haloes, we stack haloes and take the median density as a function of radius to obtain a clean density profile.
We identify the splashback feature as the region around the minimum of the logarithmic slope and bootstrap to obtain errors.
We define the depth of the feature as the height of the feature from the minimum of the logarithmic slope to where the slope flattens at a larger radius and the width of the feature at the midpoint of the depth.
We use an empirical function that models both the inner and outer halo regions of the density profile. 
We then compute the logarithmic slope of the fitted profiles and define the splashback features.
This section describes the fitting procedure, and the next section gives details about how we estimate the splashback features from the fitted logarithmic slope of the fitted profiles.
    \input{methods/methods_fit}
    
\subsection{The splashback feature}
\label{sec:Rsp}
    \input{methods/methods_features}

%% file: methods/methods_simulations.tex
In this work, we use data from the MillenniumTNG (MTNG) project \citep{hernandez2023millenniumtng, Kannan2023, Bose2023,Pakmor2023}.
MTNG builds on the Illustris \citep{Vogelsberger2014a,Vogelsberger2014b} and IllustrisTNG simulations \citep{marinacci2018first, nelson2018first, pillepich2018first, springel2018first, naiman2018first} and combines them with the much larger volume reached by the iconic dark matter-only Millennium simulation. The hydrodynamic simulations are performed using the moving-mesh code \textsc{Arepo} \citep{springel2010pur, weinberger2020arepo}.
We focus primarily on the hydrodynamic version in this work, and we briefly compare it to the dark matter-only version in Appendix \ref{apx:hydro-DM}.

Cosmological parameters used are consistent with \citet{ade2016planck}: $\Omega_m = \Omega_{\mathrm{DM}} + \Omega_\mathrm{b} = 0.3089$, $\Omega_b = 0.0486$, $\Omega_{\Lambda} = 0.6911$, $\sigma_8 = 0.8159$, $n_s = 0.9667$, and the Hubble constant $H_0=100h\ \text{km s}^{-1} \text{Mpc}^{-1}$ where $h = 0.6774$.
MTNG is a periodic cube with a side length of 500$h^{-1}$Mpc and was run at five and four resolution levels for dark matter-only and hydrodynamic simulations, respectively. For our analysis, we use the highest simulation run (MTNG-L500-4320-A). MTNG-L500-4320-A has $4320^3$ cells/particles, with a target gas cell mass of $2\times10^7h^{-1}M_\odot$ and a dark matter particle mass of $1.12\times 10^8 h^{-1}M_\odot$. The gravitational softening length is $2.5h^{-1}$ kpc. 

Halo structures are identified using the friends-of-friends (FoF) algorithm \citep{davis1985evolution}, while subhaloes are detected with the \textsc{SubFind} algorithm \citep{springel2001populating, dolag2009substructures}. 
The algorithm identifies structures by linking particles with neighbouring particles that lie within a linking length of $b = 0.2$. 
The most massive gravitationally bound object in an FoF group is the main halo and others are labelled as subhaloes. The halo centre is defined as the position of the most bound particle within the main halo by using \textsc{SubFind}.
To trace the formation histories of haloes, MTNG employs the \textsc{SubLink} merger tree algorithm, which links subhaloes across snapshots based on their most bound particle. This enables a consistent and physically motivated tracking of halo assembly over cosmic time \citep{rodriguez2015merger}.

%% file: methods/methods_halos.tex
\subsubsection{Halo mass}\label{sec:mass}

We stack haloes by mass to calculate the median density profile of the stacked set. 
We can define the size of a halo to be the radius $R_{200\text{m}}$ of a sphere with an enclosed density $\rho_{\Delta}$ of 200 times the mean matter density of the universe $\rho_\text{m}$ such that
\begin{equation}\label{eq:r200}
    R_{200\text{m}} = \left(\frac{3\,M_{200\mathrm{m}}}{4\,\pi\,\times200\rho_\text{m}}\right)^{\frac{1}{3}},
\end{equation}
and the halo mass is then $M_{200\text{m}}$, the mass enclosed within $R_{200\text{m}}$. 
When referring to a halo’s mass throughout the paper, we use $M_{200\text{m}}$.

\subsubsection{Halo mass accretion rate}

Various studies have found a relationship between $R_{\rm sp}$ and $R_{\rm st}$ and the accretion rate of a halo \citep{more2015splashback,diemer2014dependence,Diemer2017a,adhikari2014splashback,mansfield2017splashback}.
We therefore also investigate the accretion rate and use the definition from \citet{Diemer2017a} to compute the accretion rate
\begin{equation}\label{eq:gamma}
    \Gamma=\frac{\Delta \log_{10} M}{\Delta \log_{10} a},
\end{equation}
where $M$ is the halo mass and $a$ is the scale factor.  $\Delta$ refers to the change over one dynamical time, which changes with redshift $z$ and is defined as in \citet{Diemer2017a}
\begin{equation}
    t_{\text{dyn}}=\frac{t_H(z)}{5\sqrt{\Omega_m(z)}},
\end{equation}
where $t_H$ is the Hubble time $1/H(z)$.  
To compute the accretion rate of a halo at a given redshift, we first calculate the dynamical time at this redshift, then select the snapshot earlier in the simulation with a time difference closest to the calculated dynamical time. 
We note that we do not calculate the instantaneous accretion rate but the time-averaged accretion rate over the dynamical time to reduce noise from events like mergers.
We use the \textsc{SubLink} merger trees \citep{rodriguez2015merger} to identify the main progenitor of the halo. This algorithm tracks subhaloes through the simulation by linking subhaloes in consecutive snapshots that share the most bound particles. 
Therefore, by identifying the parent halo of the linked subhaloes, we can trace the history of haloes and then calculate the change in halo mass every dynamical time. 
By dividing the change in halo mass by the change in scale factor, we get the time-averaged halo accretion rate. 

\subsubsection{The most recent major merger}\label{sec:def_mergerz}
The most recent time of halo merging is important for shaping a halo's dynamical state, internal structure, and density profile \citep{Diemer2017a}.
We stack haloes by their most recent major merging time and compute the median density profile. 
We define a major merger to be a merger where the smaller halo is at least 10 percent of the larger halo's mass.
The \textsc{SubLink} merger trees \citep{rodriguez2015merger} are also used for this calculation. 
We cycle through the snapshots starting at $z=0$ and identify the main and second progenitors.
When the ratio between the two exceeds 0.1, we record the snapshot where the two haloes merged as the time of the most recent major merger.
Specifically, if the second and the main progenitor's mass ratio exceeds $0.1$ for the last time at the snapshot $n$, we choose the snapshot $n+1$ as the time when merging occurs.
It may be possible for a merger with a ratio of just under $0.1$ to have been larger than this threshold in the past but shrunk due to stripping of the smaller halo, but we maintain our method for simplicity.

\subsubsection{Concentration}\label{sec:def_conc}
The concentration of a dark matter halo characterises how centrally dense the halo is, and it is typically defined by the ratio between the outer radius of the halo, i.e. $R_{200}$, and a scale radius that marks a transition in the density profile slope \citep{Bullock2001,Wechsler2002}.
To obtain the concentration of the haloes, we define the concentration as the ratio between the scale radius $r_\mathrm{s}$ and $R_{\rm200\text{m}}$
\begin{equation}
    c = \frac{R_{\rm200m}}{r_\mathrm{s}}
    \label{eq:concentration}
\end{equation}
in which the scale radius $r_s$ is from the NFW profile \citep{Navarro1996}
\begin{equation}
    \rho(r) = \frac{\rho_\mathrm{s}}{\frac{r}{r_\mathrm{s}}\left(1+\frac{r}{r_\mathrm{s}}\right)^2}.
    \label{eq:NFW}
\end{equation}
By fitting the density profile of individual halo with the NFW profile within the range $r=0.01R_{\rm200\text{m}}$ and $r=R_{\rm200\text{m}}$, we calculate the optimal parameters $\rho_\mathrm{s}$ and $r_\mathrm{s}$, and then compute the concentration of each halo. 
Haloes are therefore stacked by concentration and the median density profile of every concentration stack can be computed. 

\subsubsection{Peak height}
The peak height of a halo relates to how overdense a halo is and has a direct correlation with halo mass and redshift \citep{press1974formation}.
The peak height $\nu$ provides a dimensionless measure of the rarity of the initial density peak from which a halo forms. It compares the critical overdensity for spherical collapse $\delta_c$, with the RMS variance $\sigma\left(M,z\right)$ of the linear density field smoothed on the mass scale $M$. 
Halos with the same peak height $\nu$ arise from equally rare peaks in the primordial density field. 
They are therefore at comparable stages of gravitational collapse and reside in statistically similar environments, independent of redshift \citep[e.g.][]{sheth1999large,diemer2015universal}.
Studies have shown that the splashback radius has an obvious correlation with peak height, so peak height can potentially correlate with other splashback features which clears out the redshift dependence \citep{more2015splashback,diemer2017splashback}.
We use \textsc{Colossus} \citep{Diemer2018} to calculate the peak height for a halo with a given mass and redshift.
The peak height $\nu$ is defined as
\begin{equation}
    \nu = \frac{\delta_c}{\sigma(M,z)}
    \label{eq:peak_height}
\end{equation}
where $\delta_c \simeq 1.686$ is the linear overdensity threshold for halo collapse, and $\sigma(M,z)$ is the RMS variance of the linear density field on a scale for mass $M$ at redshift $z$. 

\subsubsection{Formation time}\label{sec:formz_def}
There are various ways of examining the formation time of haloes, including definitions based on the redshift when the halo reaches a certain fraction of its peak mass \citep{Wechsler2002}, or the time when its main progenitor first exceeds a fixed mass threshold \citep{giocoli2010substructure}. We define the halo formation time as the redshift when the halo reaches half of its present mass $M_{200\text{m}}$. 
It is the snapshot at which the mass ratio $M_{200{\text{m}}}(z)$ and $M_{200{\text{m}}}(0)$ is closest to 0.5. 

To compute individual halo's formation time $z_{\text{form}}$, we search through the \textsc{SubLink} merger trees \citep{rodriguez2015merger} to identify the snapshot where the main progenitor of the halo has half of the $z=0$ mass.

%% file: methods/methods_samples.tex
We select samples of haloes from the hydrodynamic simulation snapshots over the redshift range $0 \leq z \leq 2$. We select an initial sample of haloes at a given redshift by taking all haloes with a total mass of $M_{200\text{m}}\geq10^{13}M_\odot$ from MTNG. We obtain 54,898 haloes at $z = 0$, 24,299 haloes at $z=1$ and 5,167 haloes at $z=2$. For the number of haloes at redshifts between 0 and 2, see Table \ref{tab:table2}.
Furthermore, since a major focus in this work is the splashback feature dependence on halo mass, we also summarise the numbers of haloes in each mass cut at every redshift in Table \ref{tab:table2}. 

We also bin haloes by other features, e.g., accretion rate. 
We use the same halo samples summarised in Table \ref{tab:table2}. Instead of binning by mass, we calculate the accretion rate of each halo at each redshift using Equation \ref{eq:gamma} and bin haloes by accretion rate. 
The same strategy is applied to any other physical quantities we investigate. 


\begin{table*}
    \addtolength{\tabcolsep}{8pt}
    \def\arraystretch{1.2}
    \centering
    \begin{tabular}{lcccccc}
        \hline
        Redshift & \multicolumn{5}{c}{Mass bins $[{\rm M}_{\odot}]$} & Total\\
         & $10^{13}\sim10^{13.5}$ & $10^{13.5}\sim10^{14}$ & $10^{14}\sim10^{14.5}$ & $10^{14.5}\sim10^{15}$ & $10^{15}\sim10^{15.5}$ & \\
         \hline
         \hline
        0.0 & 29,599 & 18,635 & 5,467 & 1,099 & 98 & 54,898 \\
        0.2 & 26,833 & 15,966 & 4,023 & 650 & 28 & 47,500 \\
        0.5 & 23,634 & 12,812 & 2,735 & 279 & 4 & 39,464 \\
        1.0 & 16,321 & 7,010 & 937 & 31 & 0 & 24,299 \\
        1.5 & 9,156 & 2,918 & 193 & 2 & 0 & 12,269 \\
        2.0 & 4,212 & 931 & 24 & 0 & 0 & 5,167 \\
        \hline
    \end{tabular}
    \caption{The number of haloes per mass cut at each redshift between $0\leq z<2$ in the MTNG hydrodynamic simulations.  The larger mass bins do not contain any haloes at higher redshift, and some of the larger mass bins contain small samples of haloes.}
    \label{tab:table2}
\end{table*}

%% file: methods/methods_fit.tex
\subsubsection{The density profiles}
We fit dark matter halo density profiles with the analytic function proposed in \citet{diemer2014dependence}
\begin{equation}
\begin{split}
&\rho(r)=\rho_{\text{inner}}\times f_{\text{trans}}+\rho_{\text{outer}} \\
&\rho_{\text{inner}}=\rho_{\text{Einasto}}=\rho_s\exp{\left(-\frac{2}{\alpha}\left[\left(\frac{r}{r_s}\right)^{\alpha}-1\right]\right)} \\
&f_{\text{trans}} = \left[1+\left(\frac{r}{r_t}\right)^{\beta}\right]^{-\frac{\gamma}{\beta}} \\
&\rho_{\text{outer}} = \rho_m\left[b_e\left(\frac{r}{5R_{\text{200m}}}\right)^{-S_e}+1\right].
\end{split}
\label{eq:profile}
\end{equation}
The parameters $\rho_s$, $r_s$, $r_t$, $\alpha$, $\beta$, $\gamma$, $b_e$ and $S_e$ are left free to vary for a given stacked density profile. This formula combines descriptions for the inner region of a halo $\rho_{\text{inner}}$, in this case an Einasto profile, and for the outer region $\rho_{\text{outer}}$, where the profile begins to flatten to the mean density of the universe. The transitional region between the inner and outer regions is described by $f_{\text{trans}}$. This formula was developed for dark matter-only simulations, and we have found that it continues to work well for hydrodynamic simulations.

We compute the profile of a halo from its centre as defined by the halo finder described in Section \ref{sec:halo_finder}. We read all dark matter particles in the simulation and bin them into 85 logarithmically spaced spherical bins between 0.01 and 5 $R_{200\text{m}}$. The density profiles are computed by summing the mass of the dark matter particles within each radial bin and dividing by the bin volume.

Individual haloes are not themselves spherically symmetric objects, and the averaged density profiles can be noisy, especially for low-mass haloes and at the outskirts of a halo. 
Therefore, we stack sets of haloes depending on the property being examined (e.g., mass or accretion rate). When grouping by mass, we separate the haloes into five logarithmically spaced mass bins, with $\log_{10}(M_{200\text{m}}/{M_\odot})$ between 13, 13.5, 14, 14.5, 15, and 15.5, respectively.
For accretion rate, we select haloes with accretion rates $0\leq\Gamma\leq 6$ and separate the haloes into 12 accretion rate bins, with $\Gamma$ between 0, 0.5, 1, 1.5, ..., 4.5, 5, 5.5, and 6. 
Before stacking haloes, we normalise the profiles by $R_{\rm200m}$ to mitigate effects from size variations smearing out the splashback feature. 
We take the median average density value in each radial bin to yield a stacked density profile as a function of median $R_{200\text{m}}$. The top panel of Figure \ref{fig:profile} is an example of density profiles of dark matter halo stacks with a mass of $10^{14}-10^{14.5} M_{\odot}$ at $z=0$. 

\subsubsection{Fitting the density profiles}

Following \citet{O'Neil2021}, we fit the slope directly by taking the numerical logarithmic derivative of the stacked halo profiles and the logarithmic derivative of Equation (\ref{eq:profile}). We compute the numerical derivative of the density profiles using a central finite difference scheme with fourth-order accuracy for the logarithmic profile
\begin{equation}
    \frac{\mathrm{d} \log\rho}{\mathrm{d} \log r}\bigg|_{r_i} =
    \frac{\frac{1}{12}\log\rho_{i-2}-\frac{2}{3}\log\rho_{i-1}+\frac{2}{3}\log\rho_{i+1}+\frac{1}{12}\log\rho_{i+2}}{\log r_{i+2}-\log r_{i-2}},
\end{equation}
\label{eq:numerical_log}
where $r_i$ is the point for which the derivative is being calculated and $r_{i\pm1}$, $r_{i\pm2}$ are the surrounding points. The logarithmic derivative of Equation \ref{eq:profile} is given by
\begin{equation}
\begin{split}
    \frac{{\rm d} \log\rho}{{\rm d} \log r}&=\frac{r}{\rho}\frac{{\rm d} \rho}{{\rm d} r} \\ 
    &=\frac{r}{\rho}\left(\frac{{\rm d} \rho_{\text{inner}}}{{\rm d}r}\times f_{\text{trans}}+\rho_{\text{inner}}\times\frac{{\rm d} f_{\text{trans}}}{{\rm d}r}+ \frac{{\rm d}\rho_{\text{outer}}}{{\rm d}r}\right),
\end{split}
\end{equation}
and explicitly, 
\begin{equation}\label{eq:gradient}
\begin{split}
    &\frac{{\rm d}\rho_{\text{inner}}}{{\rm d}r} = - \frac{2}{r_s} 
\left( \frac{r}{r_s} \right)^{\alpha - 1} 
\times \rho_{\text{inner}} \\
&\frac{{\rm d} f_{\text{trans}}}{{\rm d}r} = 
\left( 1 + \left( \frac{r}{r_t} \right)^\beta \right)^{-\frac{\gamma}{\beta} - 1}
\left( -\frac{\gamma}{r_t} \right) 
\left( \frac{r}{r_t} \right)^{\beta - 1} \\
&\frac{{\rm d}\rho_{\text{outer}}}{{\rm d}r} = 
- \frac{\rho_m b_e S_e}{5 R_{200\,\text{m}}} 
\left( \frac{r}{5 R_{200\,\text{m}}} \right)^{-S_e - 1}.
\end{split}
\end{equation}
We use the \textsc{curve\_fit} function in the \textsc{SciPy} package to fit the slope profiles.
We start with fitting the median density profile of every stack with the parameters described in Equation \ref{eq:profile}. We then use the found optimal parameters as the initial guess and fit the gradient profile with Equation \ref{eq:gradient}.
There has also been a simpler equation proposed to fit halo density profiles out to a large radius in \citet{Diemer2023}.
We explore how altering this fitting function affects our results in Appendix \ref{apx:fitting}. 

\subsubsection{Error estimation}

We use bootstrapping to estimate the uncertainty in our calculations. 
At each redshift, we randomly sample halo profiles with replacement 10000 times. 

These 10000 halo samples are then separated into bins depending on what physical quantities we are looking at. For example, when binning the 10000 haloes by mass, we separate them into the five predefined mass bins.
If there are no haloes in any bin whose original samples have haloes, we discard that sample and create another sample with 10000 haloes until all bins contain at least 1 halo.
For example, at $z=0$, all five mass bins contain haloes, so we bootstrap 10,000 haloes across all bins. At $z=2$, only three bins are populated (see Table~\ref{tab:table2}), so we ensure that the lowest three bins contain haloes with each bootstrap.
This is counted as one effective sampling. 
In this case, we move on to calculate the splashback features of the stacked haloes in each bin. 
We repeat the effective sampling 1024 times. For each bin, there are therefore 1024 splashback features calculated. 
We assume a Gaussian prior to these values and choose the median value as the splashback feature and take the $16^\text{th}$ and $84^\text{th}$ percentiles, approximating a standard deviation of a Gaussian distribution, as the errors of the splashback feature. 

%% file: methods/methods_features.tex
\begin{figure}
    \centering
    \includegraphics[width=\linewidth]{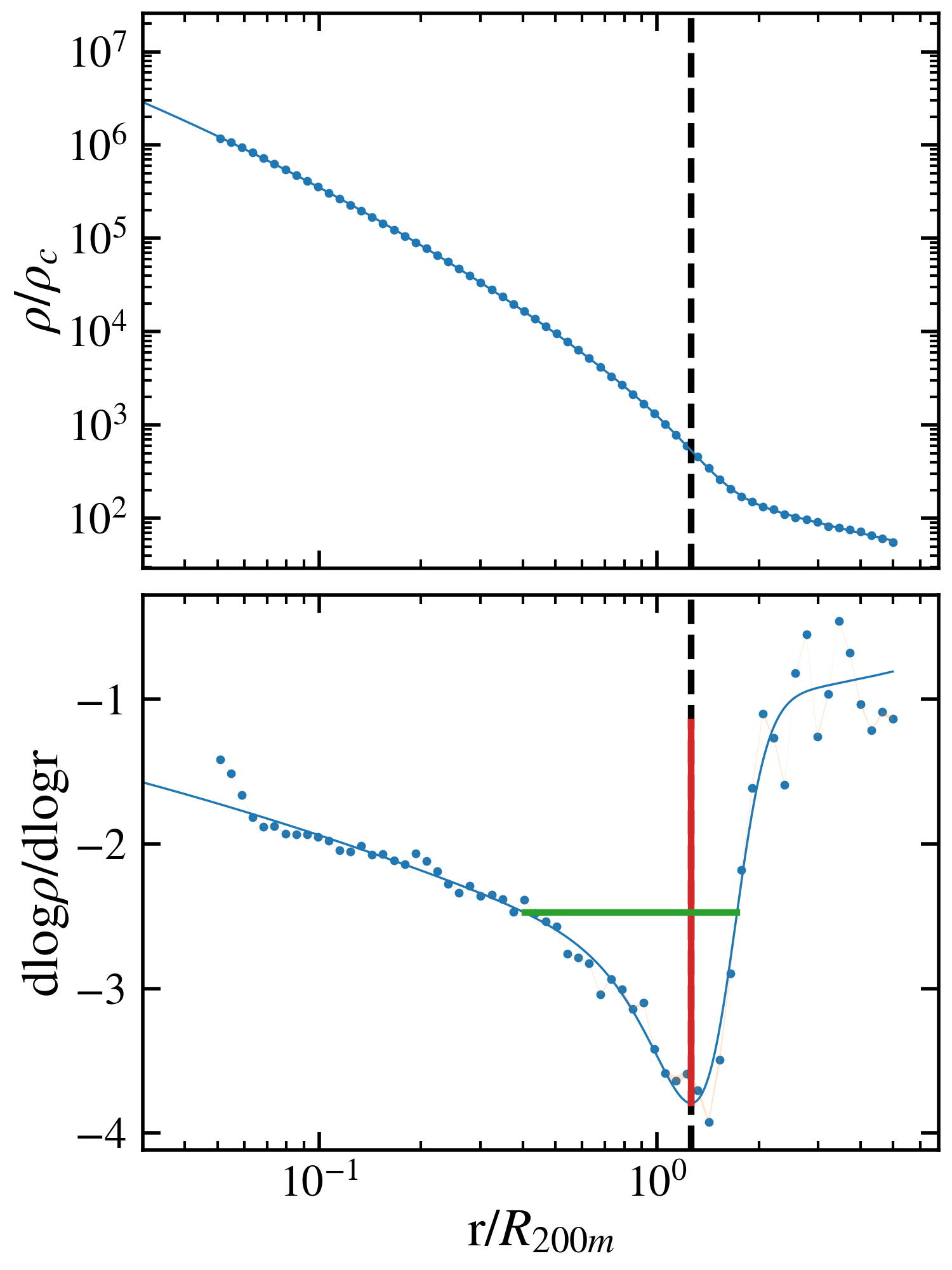}
    \caption{The upper panel is the sample density profile of the dark matter halo, and the lower panel is the logarithmic slope of the density profile. 
    The data points of the density profile come from the MTNG hydrodynamic simulations.
    This example is for the mass range  $10^{14} \leq M/\Msun<10^{14.5}M_\odot$.
    The vertical black dashed line indicates the location of the steepest slope $R_{\rm st}$ of the stacked haloes.
    The solid blue line is the fitted density profile and logarithmic slope using Equation \ref{eq:profile} and Equation \ref{eq:gradient}, respectively. 
    In the lower panel, the horizontal green and the vertical red lines indicate the width and the depth of the splashback features defined in Section \ref{sec:Rsp}.}
    \label{fig:profile}
\end{figure}

The slope of the halo density profiles has a valley shape. Apart from the position of the steepest slope, the width and depth of the logarithmic slope valley also track the kinematics at the halo boundaries. 
Therefore, in order to provide quantitative measures, we define the splashback depth $\mathcal{D}$ as the difference between the logarithmic slope at the splashback radius and the radius where the gradient profile turns over outside the halo. 
This is the point of maximum curvature in the gradient to the right of the minimum.
We compute the curvature of the slope profile in a separate procedure.
We first smooth the profile with a one-dimensional Gaussian filter and then apply standard finite-difference derivatives (using \textsc{np.gradient}) to obtain the first, second, and third derivatives. 
The point of maximum curvature is identified from the extremum of the third derivative.
Mathematically, the splashback depth is defined as
\begin{equation}\label{eq:D}
    \mathcal{D}\equiv\left.\frac{d\log\rho}{d\log r}\right|_{r_{\text{turnover}}}-\left.\frac{d\log\rho}{d\log r}\right|_{r_{\text{sp}}}.
\end{equation}
Then, the width $\mathcal{W}$ of the logarithmic slope valley is the width of the feature halfway between the height and the minimum of the feature.
A visualisation of the splashback depth and width is shown in Figure \ref{fig:profile}.
These two features relate to the splashback radius and the overall halo density profiles.
Although the depth and the width can possibly correlate with each other, we treat them as separate variables and examine how they vary with halo properties over time.
We also examine the ratio of the two in some cases to characterise the overall sharpness of the feature.

%% file: results.tex
In this section, we explore the splashback feature in a sample of haloes. We focus on the novel width, depth and depth-to-width ratio of the logarithmic slope of the halo density profile then fit them as a function of mass, redshift, accretion rate, and halo history relevant quantities.
Errors in our resulting fitting functions for depth and width are obtained from the fit covariance.
With these results, we hope to shed light on techniques to learn about a halo's less observable features and history.

\subsection{Splashback features as a function of halo mass and redshift}
\label{sec:results_mass}
\input{results/func_mass}

\subsection{Splashback features as a function of halo accretion rate and merger history}
\label{sec:results_accret}
\input{results/func_accret}
\newpage
\input{results/halo_assembly}
\label{sec:results_assembly}

%% file: results/func_mass.tex
\begin{figure*}
    \centering
    \includegraphics[width=0.45\linewidth]{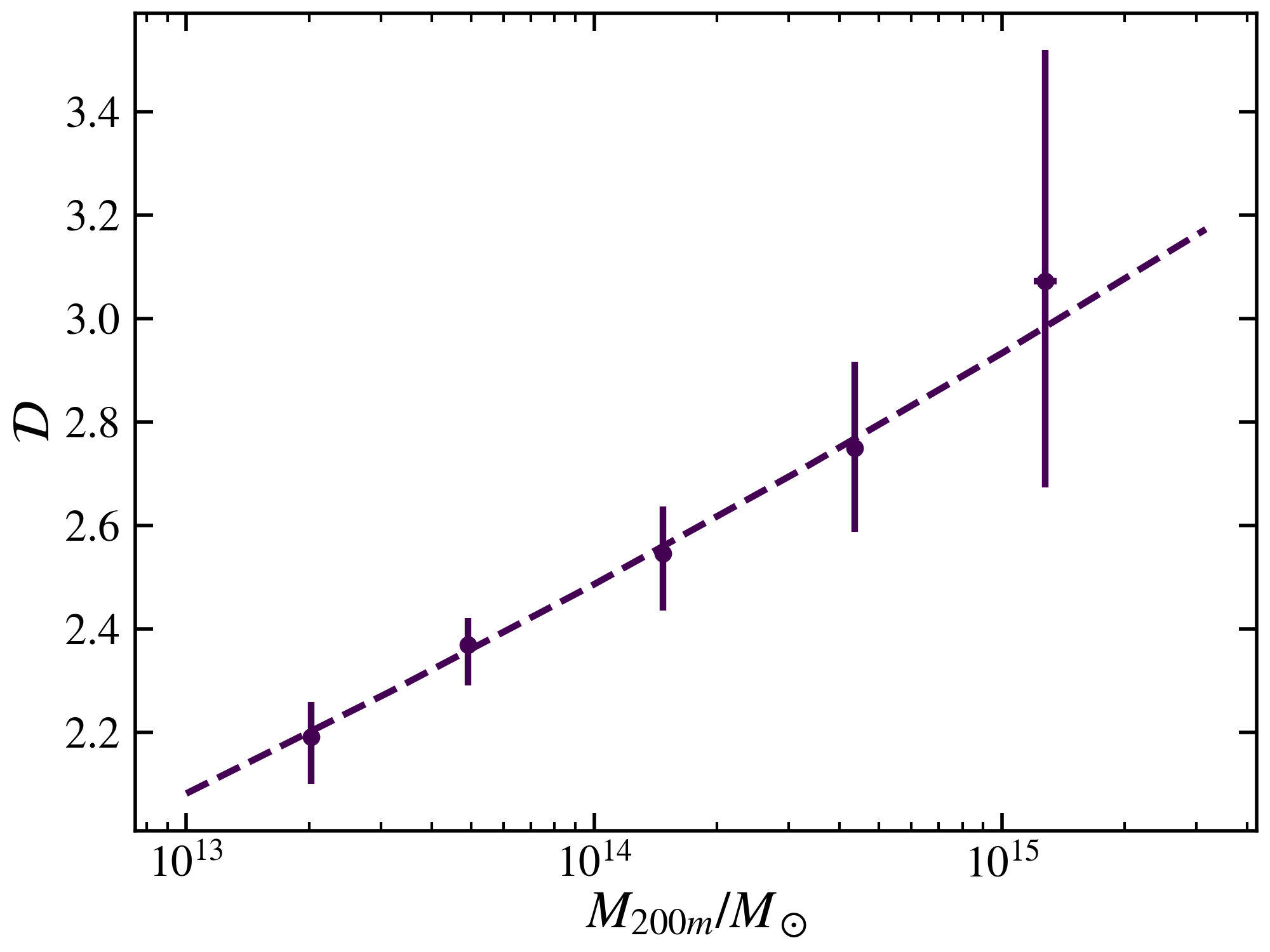}
    \includegraphics[width=0.45\linewidth]{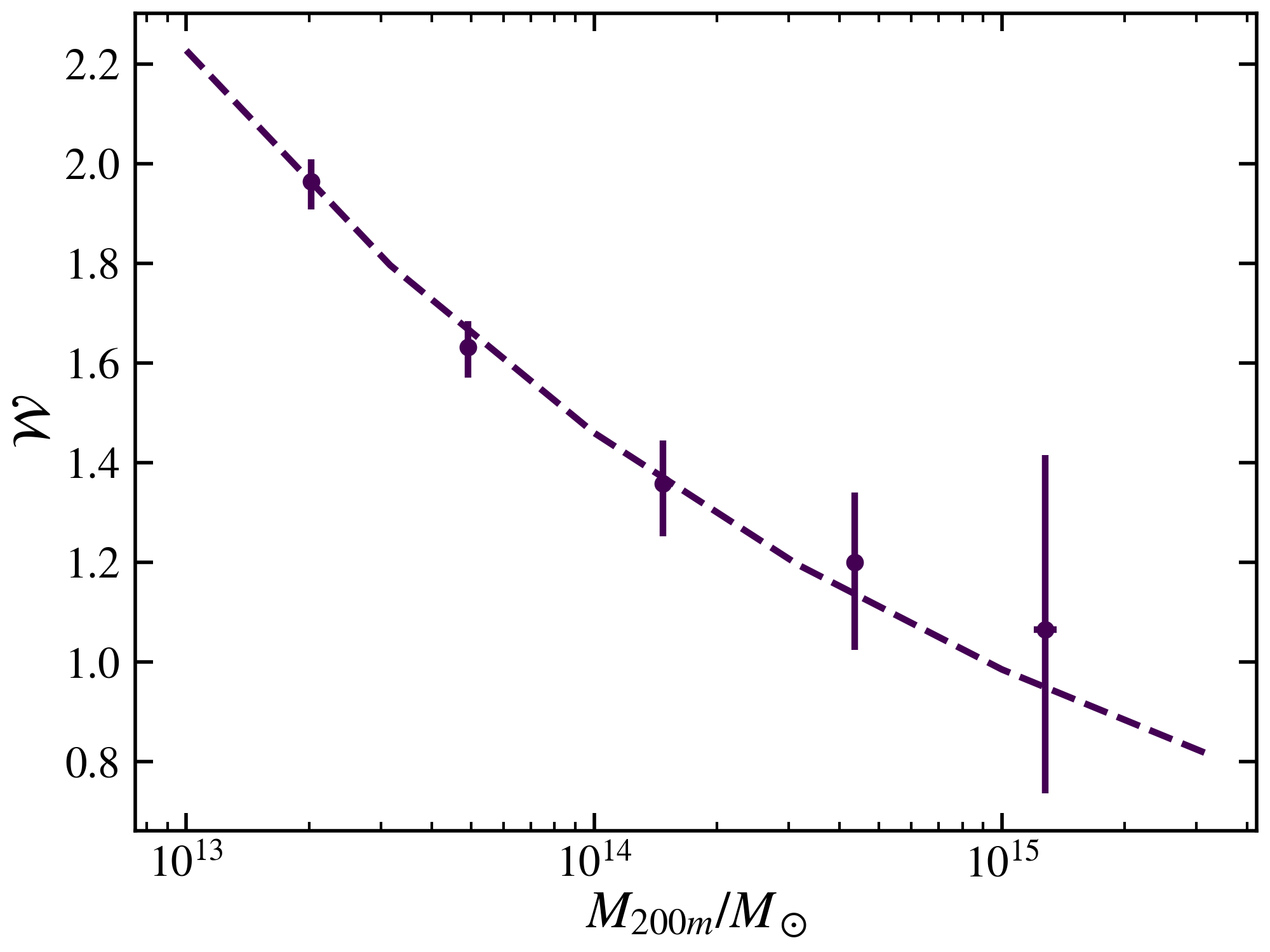}
    \caption{Depth ({\textit{Left}}) and width ({\textit{Right}}) of splashback features as a function of halo mass for the dark matter mass at $z = 0$. We stack the density profiles for haloes with $\log\left(M_{200\text{m}}/M_\odot\right)$ in 13–13.5, 13.5–14, 14–14.5, 14.5–15, and 15–15.5 and compute the splashback features of the median profile. 
    The depth increases with mass while the width decreases.  This indicates a more pronounced splashback feature for larger haloes.  Notably, this is not just an enlargement of the feature, in which both the width and depth would increase, but the decreasing width indicates a sharper decrease in the density profile.
    }
    \label{fig:mass_depth}
\end{figure*}
\begin{figure*}
    \centering
    \includegraphics[width=0.45\linewidth]{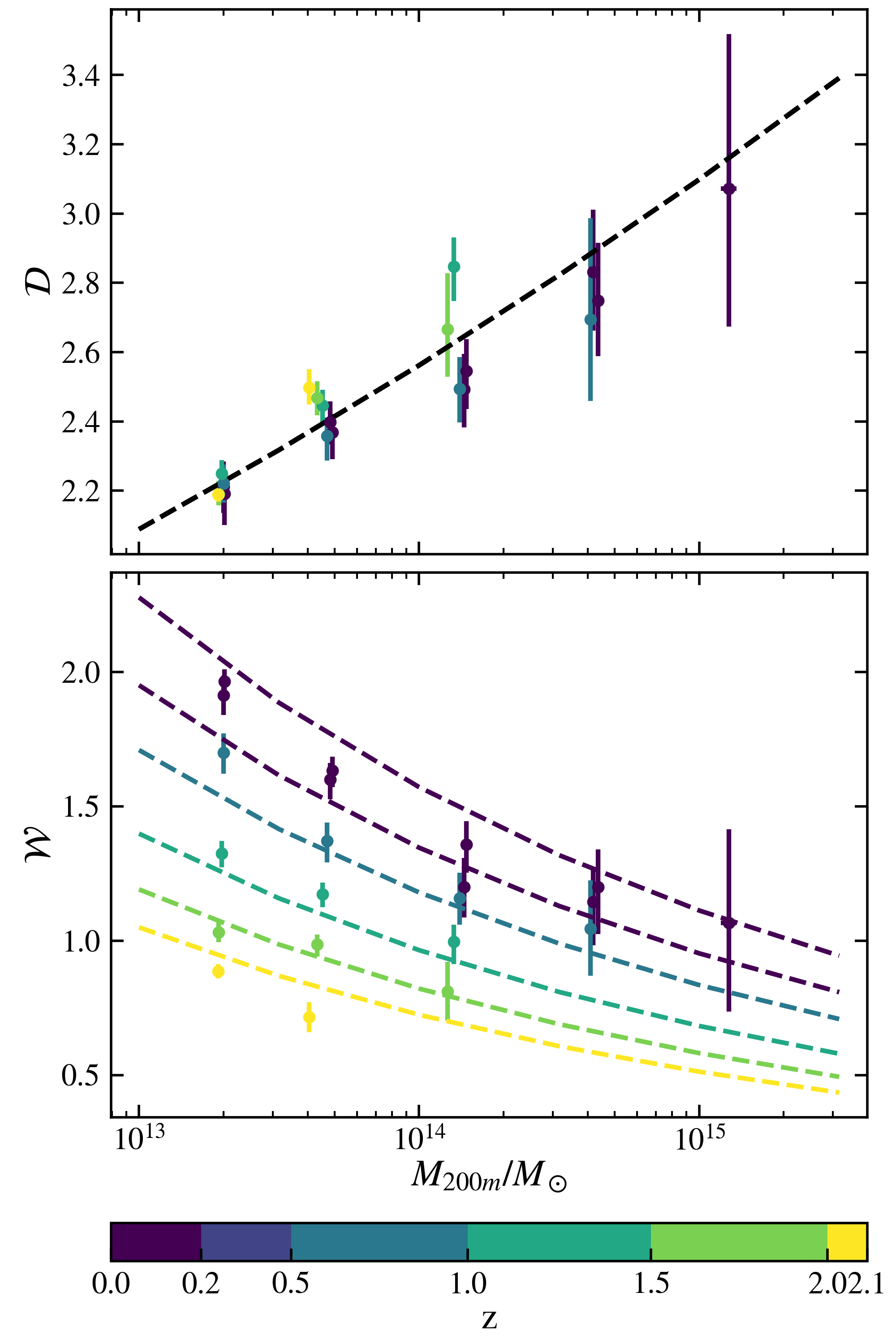}
    \includegraphics[width=0.45\linewidth]{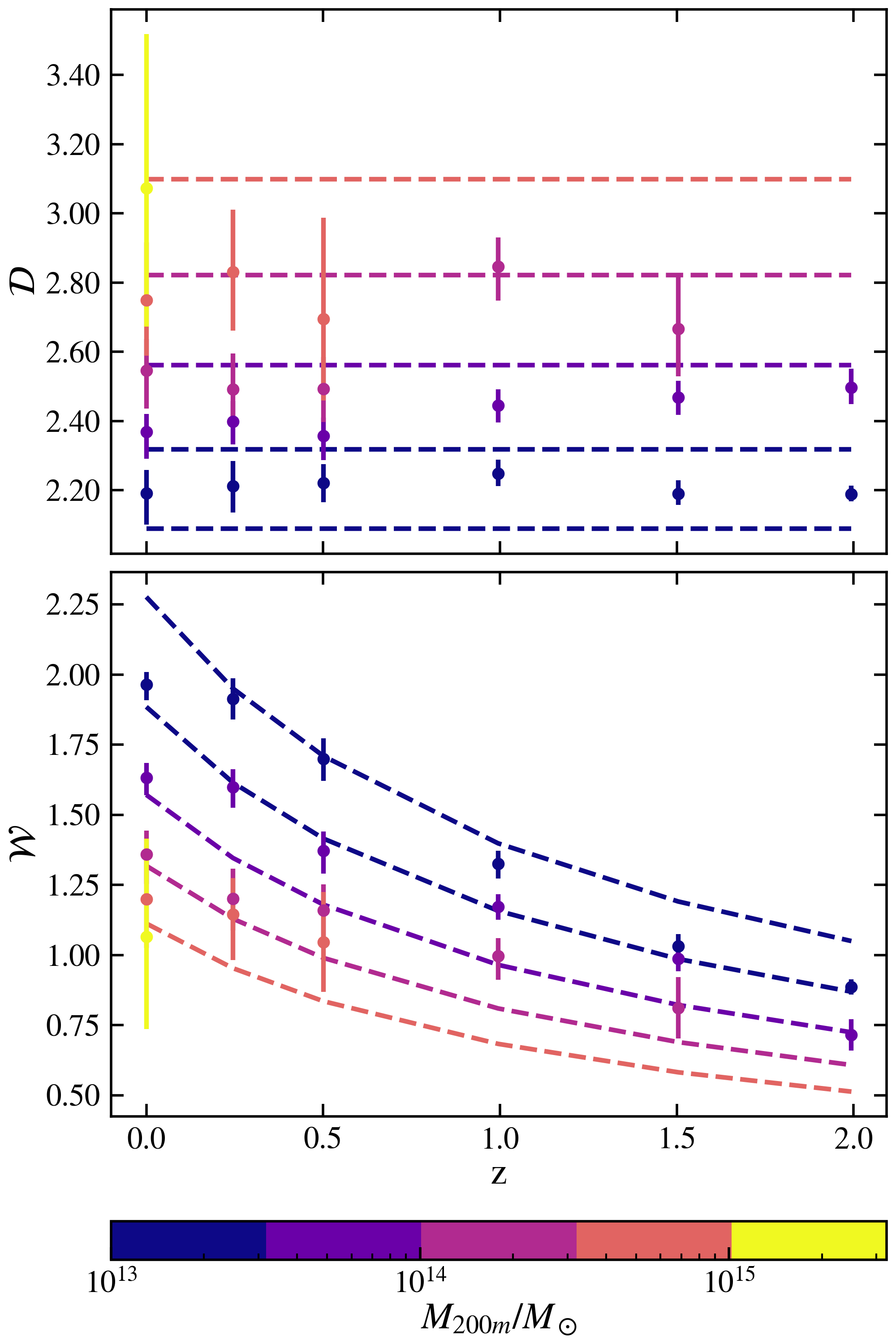}
    \caption{\textit{Top left}: Depth of splashback features as a function of halo mass for the dark matter mass at $z < 2$.  The depth linearly increases with the log of the mass.
    \textit{Top right}: Depth of splashback features as a function of redshift for the dark matter halo mass with $13\leq\log\left(M_{200\text{m}/M_\odot}\right)\leq 15.5$. 
    There is also a slight increasing trend with redshift, although this is much less pronounced than the trend with depth.
    \textit{Bottom left}: width of splashback features as a function of halo mass for the dark matter mass at $z < 2$.  The width decreases with mass, indicating that the splashback feature is narrower for larger mass haloes.
    \textit{Bottom right}: Width of splashback features as a function of redshift for the dark matter halo mass with $13\leq\log\left(M_{200\text{m}/M_\odot}\right)\leq 15.5$. 
    There is a decreasing trend with redshift, indicating that the width is more sensitive to cosmic evolution than the depth.
    The dashed lines refer to the power law fitting in Equations \ref{eq:massD} and \ref{eq:massW}.
    }
    \label{fig:mass_depth_z}
\end{figure*}

We start with examining the splashback depth and width as a function of halo mass at redshift $z=0$.
As indicated in Section \ref{sec:halo_finder}, we focus on the hydrodynamic simulation in MTNG and split the haloes into 5 mass bins between $10^{13}$ and $10^{15.5} M_\odot$. We calculate the depth and width of the splashback features of the stacked haloes in each bin by applying the fitting methods introduced in Section \ref{sec:fitting}. 
Figure \ref{fig:mass_depth} shows the change of splashback features as a function of mass at $z=0$. The depth of the splashback feature increases steadily from 2.2 to 3.0 with mass, and the width decreases from 2.0 to 1.0.
The splashback features of haloes become more significant when the halo mass is more massive.
The same computations are also performed on the dark matter-only simulation in MTNG, and the splashback features calculated are shown in Appendix \ref{apx:hydro-DM}.
There is no significant difference between the hydrodynamic and dark matter-only simulations so we will focus on the hydrodynamic simulation only in following sections.

Assuming the splashback depth and width are purely functions of mass at present, we fit each of them with a power law. 
By minimising the reduced chi-squared $\chi_\nu^2$
\begin{equation}\label{eq:chi}
    \chi_{\nu}^2 = \frac{\sum_i\left(\frac{y_{\text{obs}}-y_{\text{fit}}}{y_{\text{err}}}\right)^2}{\# \text{ of data points}-\# \text{ of params}} \; ,
\end{equation}
we find that the depth $\mathcal{D}$ of the splashback feature  ($\chi_{\nu}^2=0.041$) is given by
\begin{align}\label{eq:massD}
    \mathcal{D}(M)&=A\left(\log_{10}M\right)^B\notag \\
    &=(4.5\pm1.3)\times 10^{-3}\left[\log_{10}M\right]^{2.4\pm0.1} \; ,
\end{align}
and the width $\mathcal{W}$ ($\chi_{\nu}^2=0.19$) 
\begin{align}\label{eq:massW}
    \mathcal{W}(M)&=A\left(\log_{10}M\right)^B\notag \\
    &=(5.0\pm3.4)\times 10^6\left[\log_{10}M\right]^{-5.7\pm0.5} \; .
\end{align}
 
To generalise this expression, we examine how the splashback features change as a function of mass at different redshifts. 
We trace the redshift back to $z=2$ and calculate the splashback features at each redshift. The result is shown in the left panels of Figure \ref{fig:mass_depth_z}. 
At higher redshifts, the depth of the splashback feature consistently increases with halo mass, while its width consistently decreases. 
This trend therefore persists across redshifts. 
It indicates that more massive haloes possess clearer boundaries, and a clearer halo boundary would lead to deeper and narrower splashback features by definition of $\mathcal{D}$ and $\mathcal{W}$. 

Based on the power law fitting in Equations \ref{eq:massD} and \ref{eq:massW}, we add an extra scale factor $a=1/{(z+1)}$ dependence to the width and get the following fitting equations of depth and width of splashback features as a function of both mass and redshift (depth: $\chi_{\nu}^2=1.4$; width: $\chi_{\nu}^2=2.2$)
\begin{align}\label{eq:D_mass}
\mathcal{D}\left(M\right)=A\left(\log_{10}M\right)^B=(1.8\pm0.7)\times 10^{-3}\left(\log_{10}M\right)^{2.8\pm0.3};
\end{align}
\begin{align}\label{eq:W_mass}
    \mathcal{W}\left(M, z\right)&=A\left(\log_{10}M\right)^B(z+1)^C\notag \\
    &=(8.5\pm1.9)\times10^5\left(\log_{10}M\right)^{-5.0\pm0.6}(z+1)^{-0.71\pm0.04} \; .
\end{align}
By starting with fitting the depth with a scale factor power law dependence, we observe that $\mathcal{D}\propto (z+1)^{0.03\pm0.03}$, which eliminates the redshift dependence of the splashback depth.

Based on the existing fitting equations, we plot the splashback features predicted by halo mass and redshift, which are the dashed lines in the right panels of Figure \ref{fig:mass_depth_z}. 
For haloes with the same mass, the splashback feature is narrower in the past.
In addition, the depth does not change significantly with time for a given mass, indicating that the splashback feature was more significant in the past.
The narrower and deeper splashback features mean a more rapid change of density of halo in the splashback regions. 
Part of this trend arises from the growth of haloes: as a halo grows in size with time, by keeping the mass constant, the halo average density is smaller since particles are more spread out. The spreading of particles increases the splashback width.
Our analysis in Section \ref{sec:results_ph} further shows that, at fixed halo mass, halos at earlier times have higher peak heights $\nu$. 
Since higher $\nu$ halos originate from rarer and more spherically symmetric initial density peaks \citep{bardeen1986statistics}, it leads to a more coherent gravitational collapse, which sharpens the splashback caustic. 
The combination of these effects explains why the splashback feature is narrower and deeper at higher redshift.
The halo accretion rate also decreases with time \citep{McBride2009}. 
Slow accretion results in a more gradual transition between virialised and infalling material; thus, the splashback features are smeared out. 
This trend is further verified quantitatively in the next section through a comparison of the splashback features with different accretion rates.

Qualitatively, we observe that the depth has a smaller dependence on the redshift compared with the width, since the former is approximately constant for different redshifts and the latter is spread out with redshift.
This indicates that the splashback width is more affected by the cosmic evolution than the depth is, but overall haloes with the same mass have more distinct splashback features at higher redshifts.

%% file: results/func_accret.tex
\begin{figure}
    \centering
    \includegraphics[width=\linewidth]{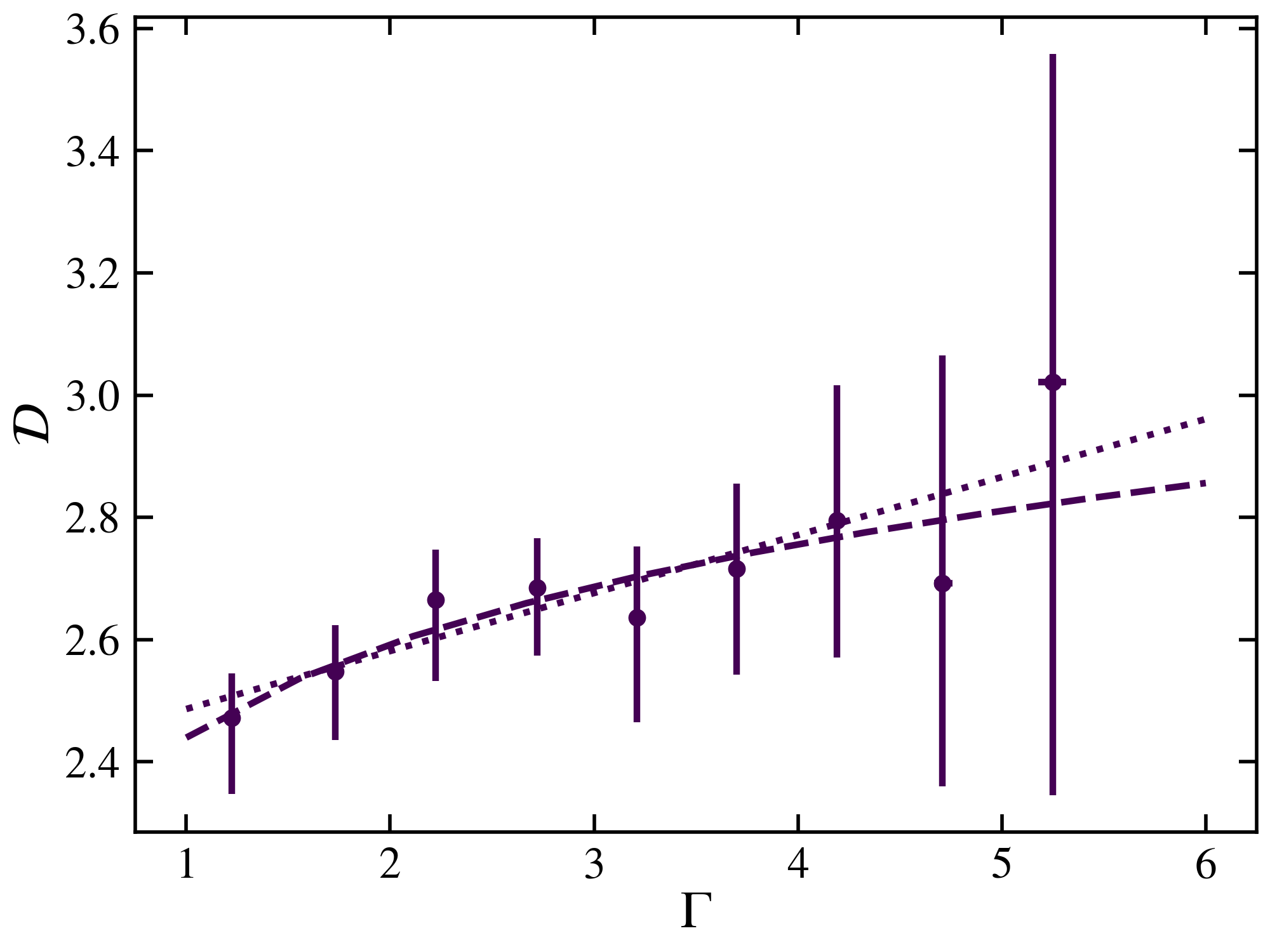}
    \includegraphics[width=\linewidth]{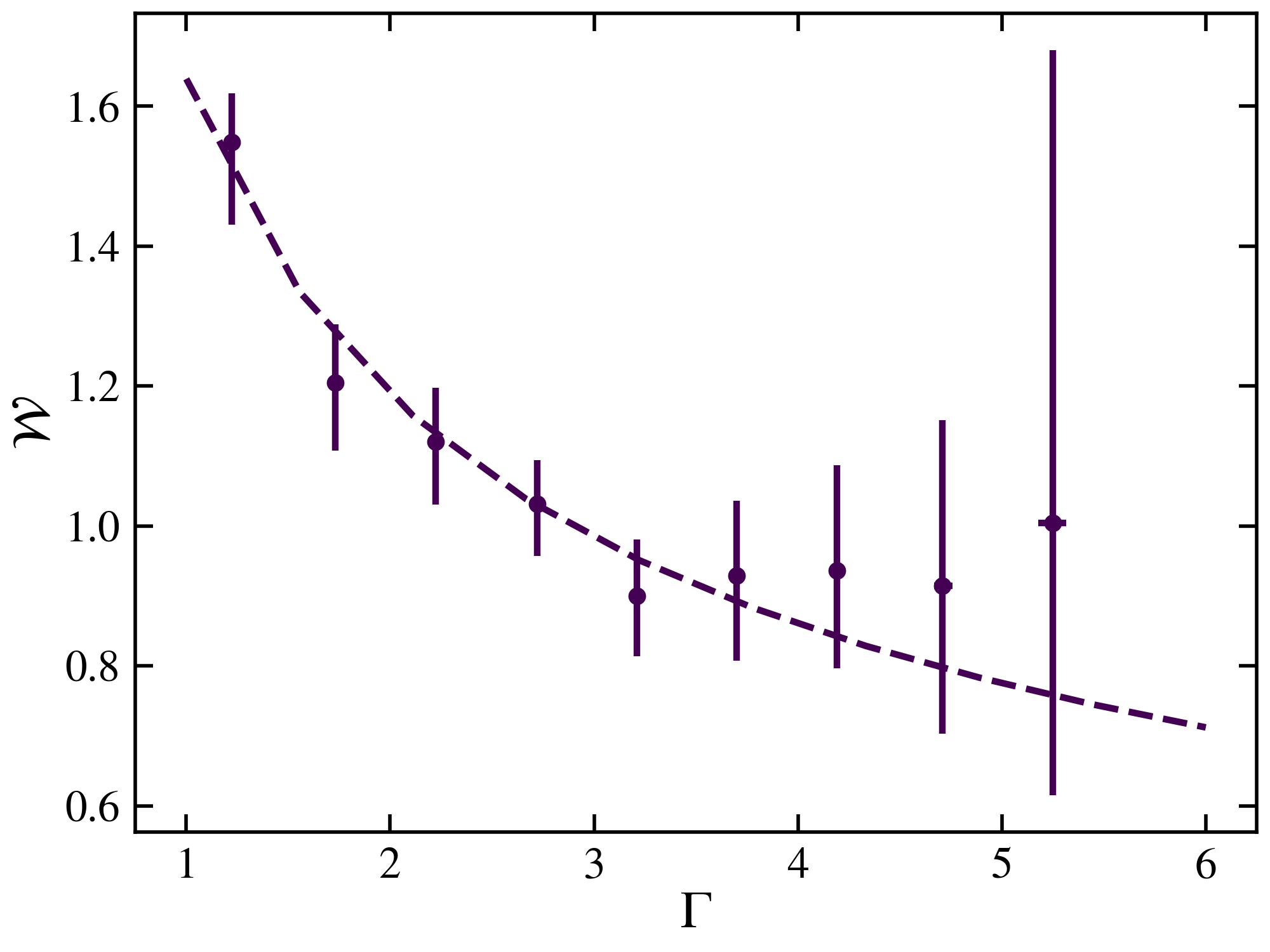}
    \caption{Depth ({\textit{Top}}) and width ({\textit{Bottom}}) of splashback features as a function of halo accretion rate for the dark matter mass at z = 0.
    The points show the data from the simulation. The dashed line represents the power law fitting while the dotted line in the upper panel represents the linear fit.
    The depth increases with accretion rate while the width decreases, making the splashback feature significantly narrower at higher accretion rates.
    }
    \label{fig:acc_depth}
\end{figure}
\begin{figure*}
    \centering
    \includegraphics[width=0.45\linewidth]{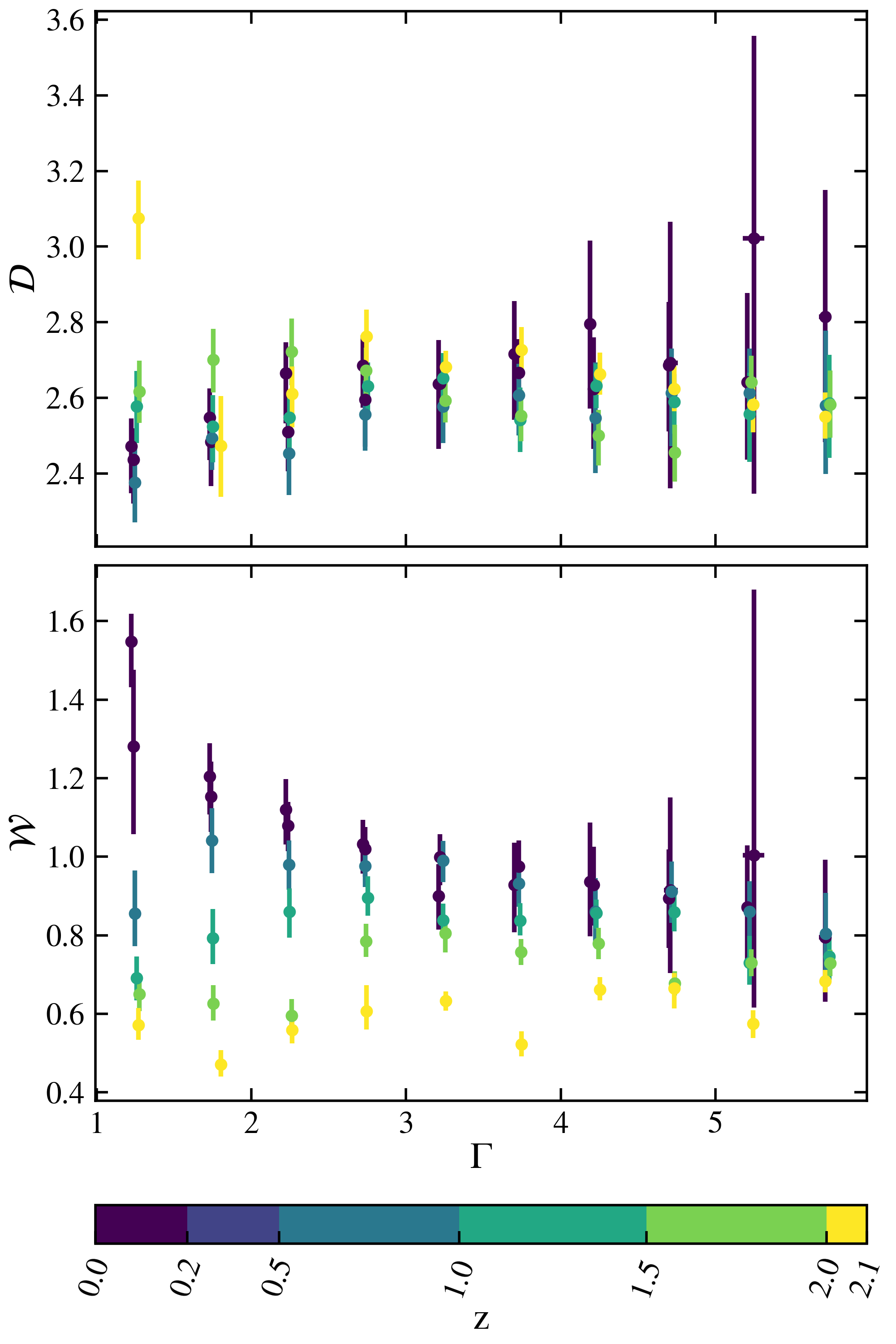}
    \includegraphics[width=0.45\linewidth]{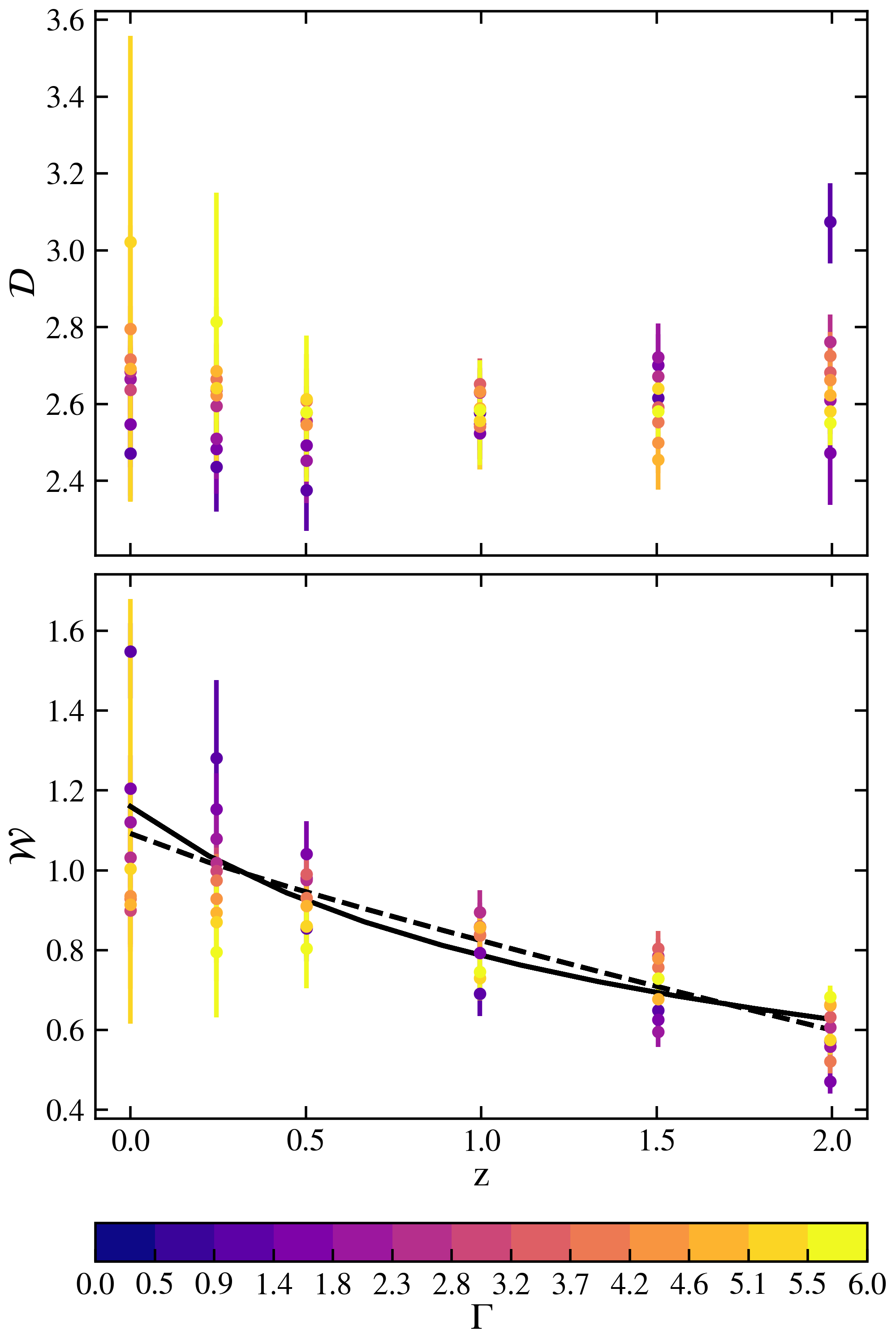}
    \caption{Relationships between width ($\mathcal{W}$) and depth ($\mathcal{D}$) of the splashback feature as a function of redshift ($z$) and accretion rate ($\Gamma$). When haloes are stacked by accretion rate, the width increases with time. However, the depth is uncorrelated with redshift and both width and depth appear only weakly dependent on accretion rate.
    In all panels, points show the data from fits in the simulation, with colour indicating either redshift (left) or accretion rate (right). 
    Haloes are stacked into accretion rate bins between $\Gamma=0$ and $\Gamma=6$ and the median profile is fitted.
    \textit{Top left}: Depth of the splashback features as a function of accretion rate for various redshifts.
    \textit{Top right}: Depth of the splashback feature as a function of redshift for various accretion rates. 
    There is no clear trend of the depth and width as a function of accretion rate.
    \textit{Bottom left}: Width of the splashback features as a function of accretion rate for various redshifts.
    \textit{Bottom right}: Width of the splashback feature as a function of redshift for various accretion rates.
    The black dashed line is fitted with the redshift only, while the black solid line is fitted using the scale factor.
    There is not much of a trend for the depth and width in terms of halo accretion rate, so we fit a function only in terms of redshift.
    There is less dependence on redshift when separated by accretion rate compared to the mass bins, as shown in Figure \ref{fig:mass_depth_z}, although there is still some dependence, with the depth decreasing significantly between redshifts $0<z<1$ and with the width peaking at redshift $z\sim1$.
    }
    \label{fig:accret_depth}
\end{figure*}
\begin{figure*}
    \centering
    \includegraphics[width=0.45\linewidth]{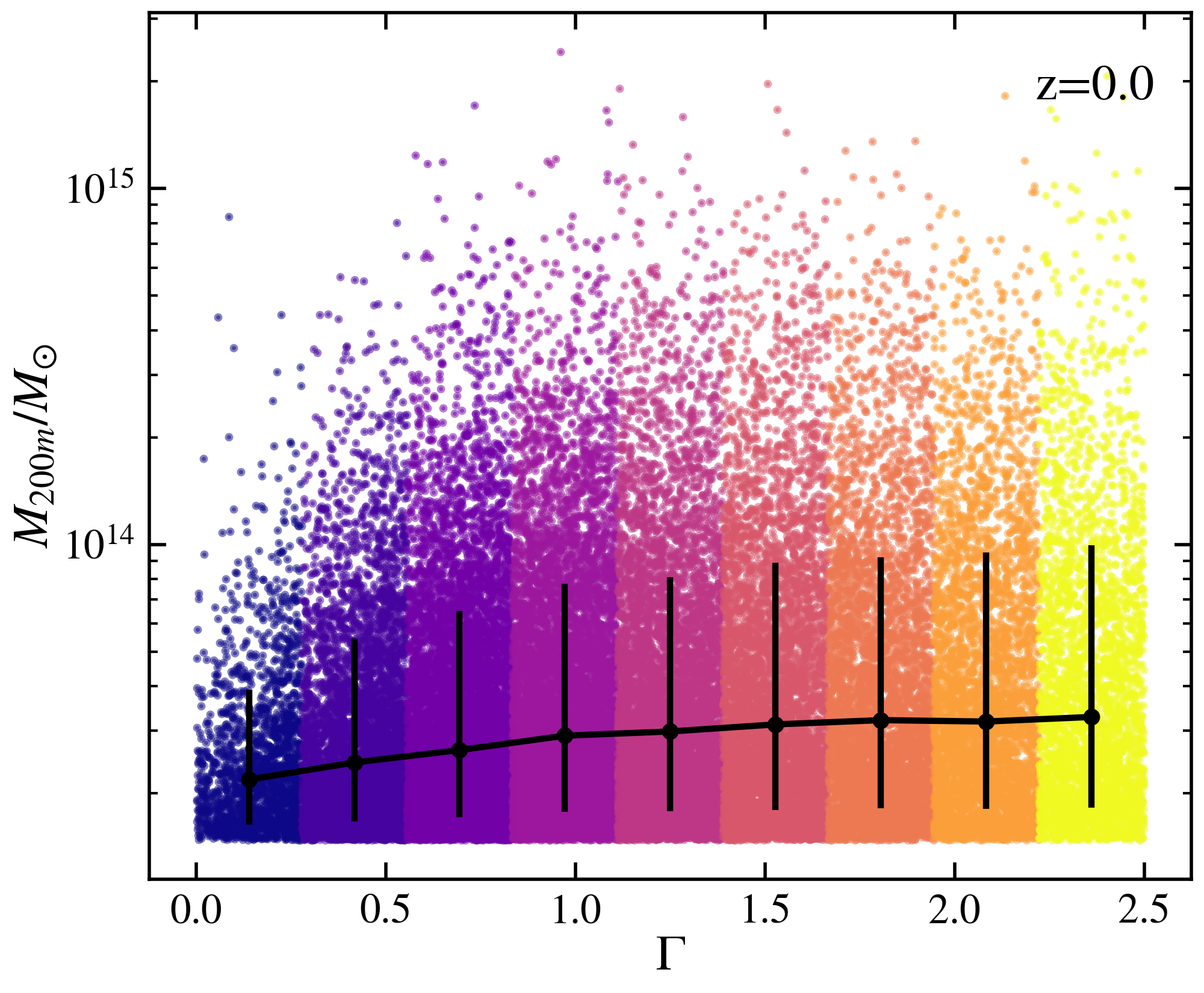}
    \includegraphics[width=0.45\linewidth]{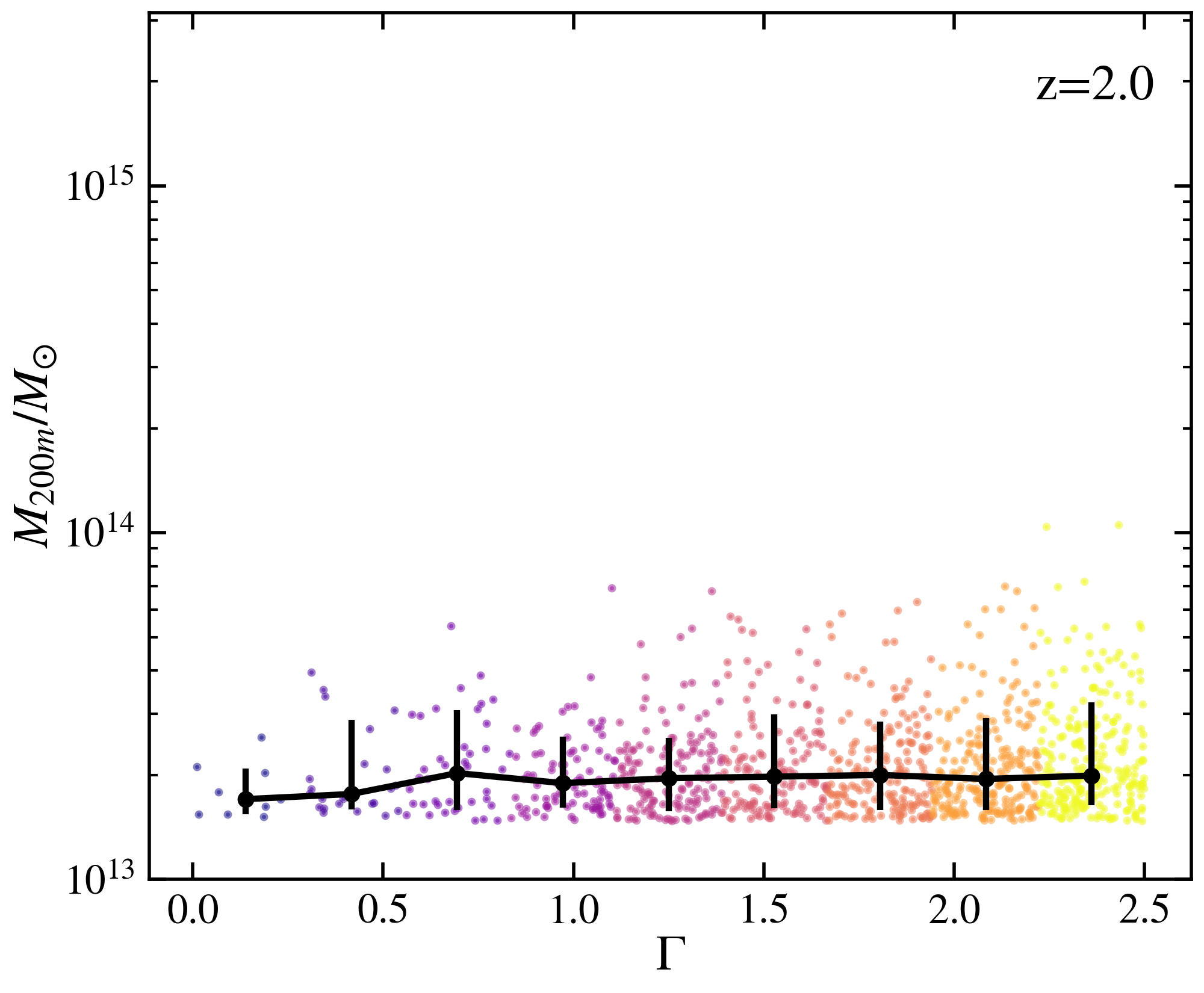}
    \caption{
    The median mass in each bin has a slight correlation with accretion rate at $z=0$ and has little correlation at $z=2$. We show the mass accretion rate relations at $z=0$ (left) and $z=2$ (right). The black line indicates the median mass in each accretion bin. 
    The colours correspond to the accretion rate bins used in the previous fitting for visualisation. 
    }
    \label{fig:acc_mass}
\end{figure*}
The assembly history of haloes, including the timing and nature of mergers, influences the splashback radius by disrupting the orbits of infalling material.
Several studies \citep[e.g.][]{diemer2014dependence,adhikari2014splashback} show that the splashback feature has a strong dependence on the halo mass accretion rate. \citet{diemer2014dependence} notes that high accretion rate haloes with $\Gamma\gtrsim3$ have a logarithmic slope that reaches approximately $-4$ while lower accretion rate haloes with $\Gamma\lesssim1$ have logarithmic slopes that reach only $-3$. This observation correlates with the splashback depth defined in our study.

We capture the dependence of the depth and width on the halo accretion rate defined in Equation \ref{eq:gamma}. 
We select haloes with accretion rates less than $\Gamma=6$ and and stack the haloes into bins with an interval of $\Delta \Gamma=0.5$. 

We start with haloes at $z=0$.
In Figure \ref{fig:acc_depth}, the splashback depth gradually increases from 2.4 to 3.0 with the accretion rate from $\Gamma=0$ to $\Gamma=5$. 
The result matches the observations and conclusions in \citet{diemer2014dependence}.
Although the definition of the splashback depth is not the same as the logarithmic slope minimum in \citet{diemer2014dependence}, they demonstrate a similar trend. 
We observe that, at $\Gamma\approx1$, the splashback depth is approximately $\mathcal{D}=2.4\pm0.1$; at $\Gamma\approx3$, the splashback depth is around $\mathcal{D}=2.7\pm0.1$. This value is consistent with that found in \citet{diemer2014dependence}.
Since \citet{diemer2014dependence} did not provide measurements beyond $\Gamma\geq 3$, we extrapolate their reported trend to $\Gamma=5$. Our simulations yield a splashback depth of $3.0\pm0.6$, consistent with this extrapolation.
We attempt to apply the numerical fitting to the splashback depth $\mathcal{D}$ in terms of halo accretion rate.
We implement a power law and a linear fitting as below
\begin{align}
    \mathcal{D}_{\text{power}}(\Gamma)=A\Gamma^B=(2.4\pm0.02)\Gamma^{0.088\pm0.01};
    \label{eq:D_acc_z0_pow}
\end{align}
\begin{align}
    \mathcal{D}_{\text{linear}}(\Gamma)=A\Gamma+B=(0.095\pm0.018)\Gamma+{(2.4\pm0.05)},
    \label{eq:D_acc_z0_lin}
\end{align}
and they respectively possess reduced $\chi_{\nu}^2=0.10$ and $\chi_{\nu}^2=0.15$. 
It is unclear which fitting method better describes the data. 
In the following section, we show that the splashback depth scales as a power law with peak height.
One possible explanation for this scaling is its indirect connection to the halo accretion rate. 
As shown in \citet{diemer2014dependence}, the accretion rate increases approximately linearly with peak height, suggesting that the depth may inherit this dependence through the accretion history of the halo.

The splashback depth increases with halo accretion rate while the width decreases. 
A sudden decrease of width from $\mathcal{W}=1.6$ to $\mathcal{W}=0.9$ occurs between $1\leq\Gamma\leq3$, and then the splashback width tends to stabilise at $\Gamma>3$.
The trend is more prominent with the splashback width than the splashback depth. 
We applied the fitting of the width $\mathcal{W}$ and in Figure \ref{fig:acc_depth}.
It shows that, at $z=0$, 
\begin{align}
    \mathcal{W}_{\text{power}}(\Gamma)=A\Gamma^B=(1.6\pm0.1)\Gamma^{-0.47\pm0.07}
    \label{eq:W_acc_z0}
\end{align}
with $\chi_{\nu}^2=0.33$.
This indicates that
\begin{equation}
    \mathcal{W}_{\text{power}}(\Gamma)\propto\frac{1}{\sqrt{\Gamma}} \;.
\end{equation}
Therefore, faster accreted haloes have more pronounced halo boundaries and more distinct splashback features. 
The decrease of the splashback width is also correlated with the results found in Figure \ref{fig:mass_depth}. 
Since the halo accretion rate $\Gamma$ increases with the halo mass $M_{200\text{m}}$ and peak height, the power law decrease of the splashback width with accretion rate is consistent with that found in Section \ref{sec:mass} and also the following Section \ref{sec:results_ph}. 
As haloes accrete matter more rapidly, newly infalling particles pile up more tightly near the first apocentre, leading to a steeper density gradient and a narrower transition zone between the virialised and infalling regions. This results in a narrower splashback width and a more pronounced splashback feature.

In order to check if the splashback feature as a function of accretion rate also persists at higher redshifts, we go back to $z=2$ and plot the depth and width in terms of accretion rate and redshift in Figure \ref{fig:accret_depth}.
At higher redshift, the splashback features tend to be more independent of the halo accretion rate and the previous power law fitting fails.
The depth varies between $\mathcal{D}=2.4\sim2.8$, and the width remains constant between $1\leq\Gamma\leq 6$.
At $z\sim2$ and $\Gamma\gtrsim4$, the splashback depth decreases with accretion rate. 
By examining the halo profiles at $z=2$, this can be explained because faster accreting haloes have more turbulent environments (i.e., more density fluctuations at the edges) \citep{klessen2010accretion}, which causes the outskirts past $R_{\rm st}$ to flatten to a lower value at a large radius.
Meanwhile, the width at higher redshifts remains insensitive to accretion rate, indicating that the halo boundary remains largely unaffected by accretion rate. Overall, the accretion rate at high redshift affects the halo environment more than the halo internal density structure. 

To account for this difference, we examine the halo mass-accretion rate distributions at $z=0$ and $z=2$, which are shown in Figure \ref{fig:acc_mass}. 
It shows that, at present, among different accretion bins, the halo mass distributions are different. 
At $z=0$, the median halo mass in each accretion bin is weakly correlated with accretion rate.
In contrast, at $z=2$, the median halo mass remains approximately constant across accretion bins.
This may be partially due to the lack of large haloes at $z=2$ making the trend less significant at higher redshift.
Therefore, when bootstrapping haloes in each accretion rate bin, there is a stronger dependence on the halo mass at $z=0$ and so higher similarity with splashback features binned by mass in Figure \ref{fig:mass_depth_z}. On the other hand, as the mass distributions in each accretion bin are similar at $z=2$, this removes the mass dependence during bootstrapping. 

There is a stronger trend as a function of redshift when binning haloes by accretion rate compared to binning by mass. Therefore, we fit the halo splashback features as a function of redshift.
We start with the redshift fitting 
\begin{equation}
    \mathcal{W}_{z}(z)=Az^B+C=\left(-0.27\pm0.07\right)z^{0.87\pm0.22}+\left(1.1\pm0.06\right)
    \label{eq:W_acc_z}
\end{equation}
with $\chi_{\nu}^2=2.4$.
The constant $C$ is the splashback width at $z=0$. More specifically, it is a halo mass accretion rate-dependent variable $C(\Gamma)$ at the present, which is effectively Equations \ref{eq:D_acc_z0_pow} and \ref{eq:W_acc_z0}.
The width can therefore be fitted as
\begin{align}
    \mathcal{W}_{z}(z,\Gamma)&=Az^B+\mathcal{W}(\Gamma)\notag \\
    &=\left(-0.28\pm0.07\right)z^{0.84\pm0.20}+\left(1.1\pm0.1\right)\Gamma^{0.034\pm0.018}
    \label{eq:W_acc_z_pow}
\end{align}
with $\chi_{\nu}^2=2.3$.
The fitting coefficient indicates that the splashback width is linearly dependent on redshift.
By comparing two sets of equations, we can observe that the fitting parameters match with each other. 
 
Another form of the redshift fitting equation is a scale factor power law which is implemented in Section \ref{sec:results_mass}. We also fit the width in terms of scale factor (width: $\chi_{\nu}^2=2.7$)
\begin{align}\label{eq:W_acc_a}
    \mathcal{W}_{a}(z)&=A(z+1)^B=\left(1.16\pm0.04\right)z^{-0.56\pm0.04}.
\end{align}
The constant $A$ in the fitting equations is regarded as the width at $z=0$ (from Equations \ref{eq:D_acc_z0_pow} and \ref{eq:W_acc_z0}). Therefore, a more physical fitting is 
\begin{align}
    \mathcal{W}_{a}(z)&=\mathcal{W}_{\text{power}}(\Gamma)(z+1)^B\notag\\
    &=(1.12\pm0.06)\Gamma^{0.035\pm0.030}(z+1)^{-0.56\pm0.04}\;
\end{align}
with $\chi_{\nu}^2=2.7$.  $\mathcal{W}_{\rm power}$ is defined in Equation \ref{eq:W_acc_z0}.
Both fitting methods develop similar fitting accuracies.
They become equivalent at small redshifts but behave differently at larger redshifts.
At this level, only more data points or reduced error bars can improve the fitting. 
Alternative simulations can be tested with the above fitting equations.

Overall, the accretion rate is an instantaneous property of halo while the mass is a cumulative historical feature of halo.
Although here we use the accretion rate averaged over a dynamical time, and it is therefore not entirely instantaneous, the mass is accumulated over the entire history of the halo.
The splashback depth and width have a stronger dependence on the long term factors, such as mass, and they are less sensitive to the shorter term factors, such as the accretion rate.
In order to verify this, we examine the splashback feature dependence on the halo relaxation state which is another short term factor. 

\begin{figure*}
    \centering
     \includegraphics[width=1\linewidth]{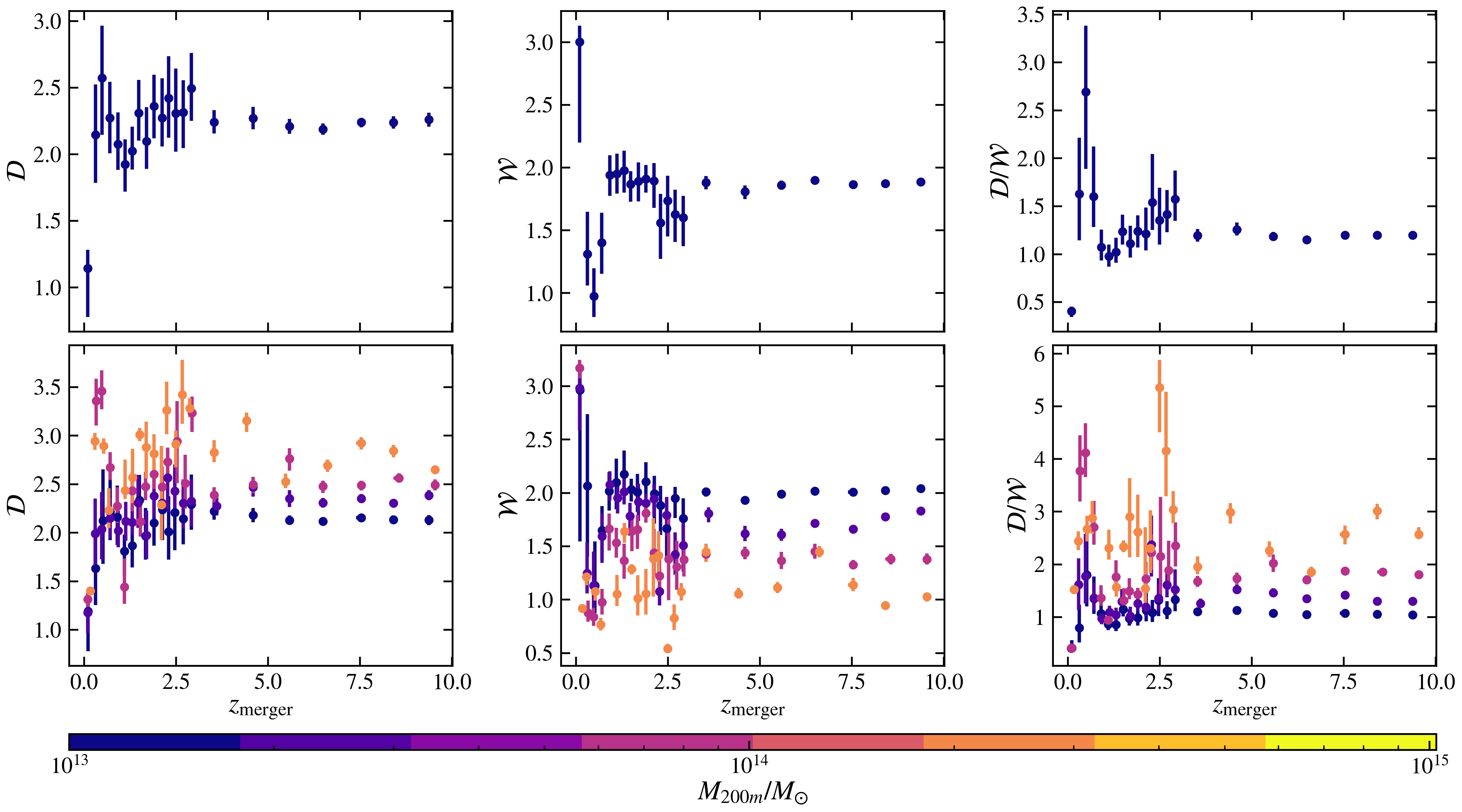}
    \caption{The splashback features have little dependence on the halo merging time.
    The most recent major merger time is defined as the time when the halo experienced a merger with a mass ratio of at least 0.1.  
    We calculate the most recent major merger time only for haloes at $z=0$.
    \textit{Left:} Depth of splashback features as a function of halo merger time. 
    \textit{Middle:} Width of splashback features as a function of halo merger time. 
    \textit{Right}: The ratio of depth and width as a function of halo merger time.
    In the bottom panels, we also bin by halo mass at $z=0$.
    }
    \label{fig:mergerz_submass_bin}
\end{figure*}

Unrelaxed haloes that are still adjusting post-merger may exhibit disrupted or less refined splashback features due to chaotic particle orbits compared to relaxed haloes with more stable structures. \citet{more2016detection} noted that the variation in $\frac{R_{\rm sp}}{R_{\rm200m}}$ due to the variation in relaxation state is approximately 0.2.
\citet{Wang2022} also showed that mergers can reduce the splashback radius by up to $10-15\%$ in individual haloes.
Therefore, we define the halo merging time in Section \ref{sec:def_mergerz} to track this effect. 
The halo merges when the ratio between the halo's second most massive subhalo and the most massive subhalo is greater than 0.1.
We focus on the haloes at redshift $z=0$. 
We select haloes with a major merging time less than $z=9$ and and stack the haloes into bins with an interval of $\Delta z=0.5$.
In the upper panel of Figure \ref{fig:mergerz_submass_bin}, the splashback depth increases before $z=3$, and tends to stability afterwards. 
This verifies that relaxed haloes which merged at higher redshift tend to have more stable splashback features compared with unrelaxed haloes which merge at $z<3$.
The splashback width is relatively constant at $z<6$ due to large error bars. At $z>6$, the splashback width reduces slightly for $\Delta \mathcal{W}\approx0.1$.

However, the overall density profile is the same for haloes with both recent major mergers and haloes with major mergers that occurred longer ago. 
This observation is also consistent with the finding of \citet{diemer2014dependence}, which showed that recent mergers do not alter the stacked halo density profiles when binned by accretion rate.
Therefore, the splashback feature depth-to-width ratio is independent of the halo merging time.
One thing to note is that these are stacked profiles, but the mergers occur in individual haloes and at only one point on the halo's edge.
Thus, the effects of mergers could be washed out by stacking, which by design removes spherical asymmetries in the halo profiles.
It may be possible to align the stacked haloes by merger direction or stack individual objects by solid angle to better capture these effects, but we leave this for future work.

The splashback features demonstrate less dependence on accretion rate and merging time which are short term factors. 
This shows that instead of being similar to the splashback radius which has a stronger correlation with halo accretion rates, the splashback depth and width depend more on the long term features of haloes. 
This is possibly due to the depth and width being measures of halo density profile rather than the tracker of particles' first turnaround at the halo boundary. 

%% file: results/halo_assembly.tex
\subsection{Splashback feature as a function of halo formation history}

Haloes are influenced by their formation history, including when and where a halo was formed. \citet{Shin2023} noted that older haloes show a higher value for $R_{\rm sp}$ by $\sim3-5\%$ when controlled for accretion rate.
Therefore, we investigate the splashback features dependence on the halo peak height, concentration as well as halo formation time respectively. 

\subsubsection{Concentration}
\label{sec:results_conc}
\input{results/conc}
\begin{figure}
    \centering
    \includegraphics[width=\linewidth]{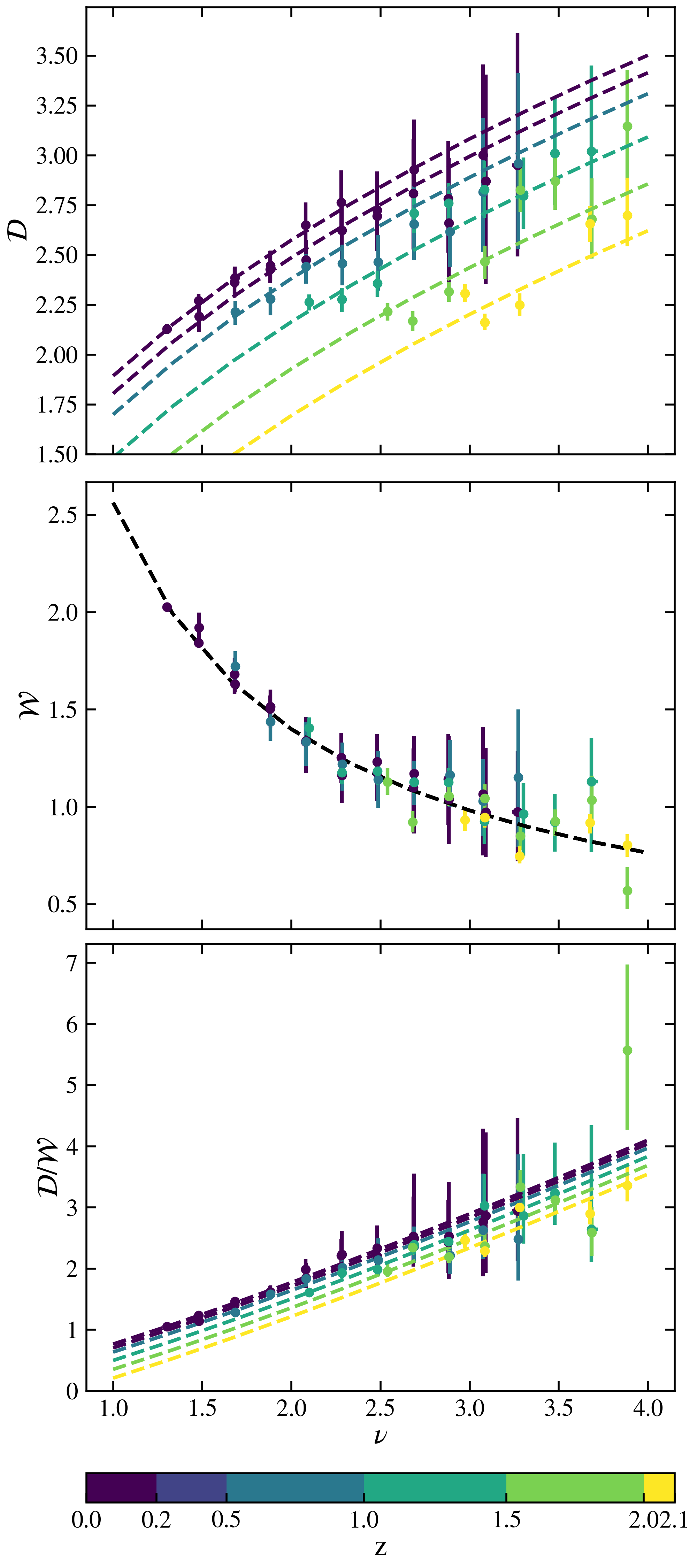}
    \caption{Haloes are stacked into peak height bins between $\nu=0$ and $\nu=5$. 
    \textit{Top:} Depth of splashback features as a function of halo peak height with colours representing redshift.
    The splashback depth has a scattered dependence on the peak height which indicates that the splashback depth binned by peak height intrinsically carries more redshift dependence.
    \textit{Middle}: Width of splashback features as a function of peak height with colours representing redshift. 
    The splashback width demonstrates a consistent dependence on peak height with little dependence on redshift.
    \textit{Bottom}: Depth-to-width ratio of splashback features as a function of peak height with colours representing redshift. 
    The ratio increases with peak height with little dependence on redshift.
   }
    \label{fig:feature_peakheight}
\end{figure}
\begin{figure}
    \centering
    \includegraphics[width=\linewidth]{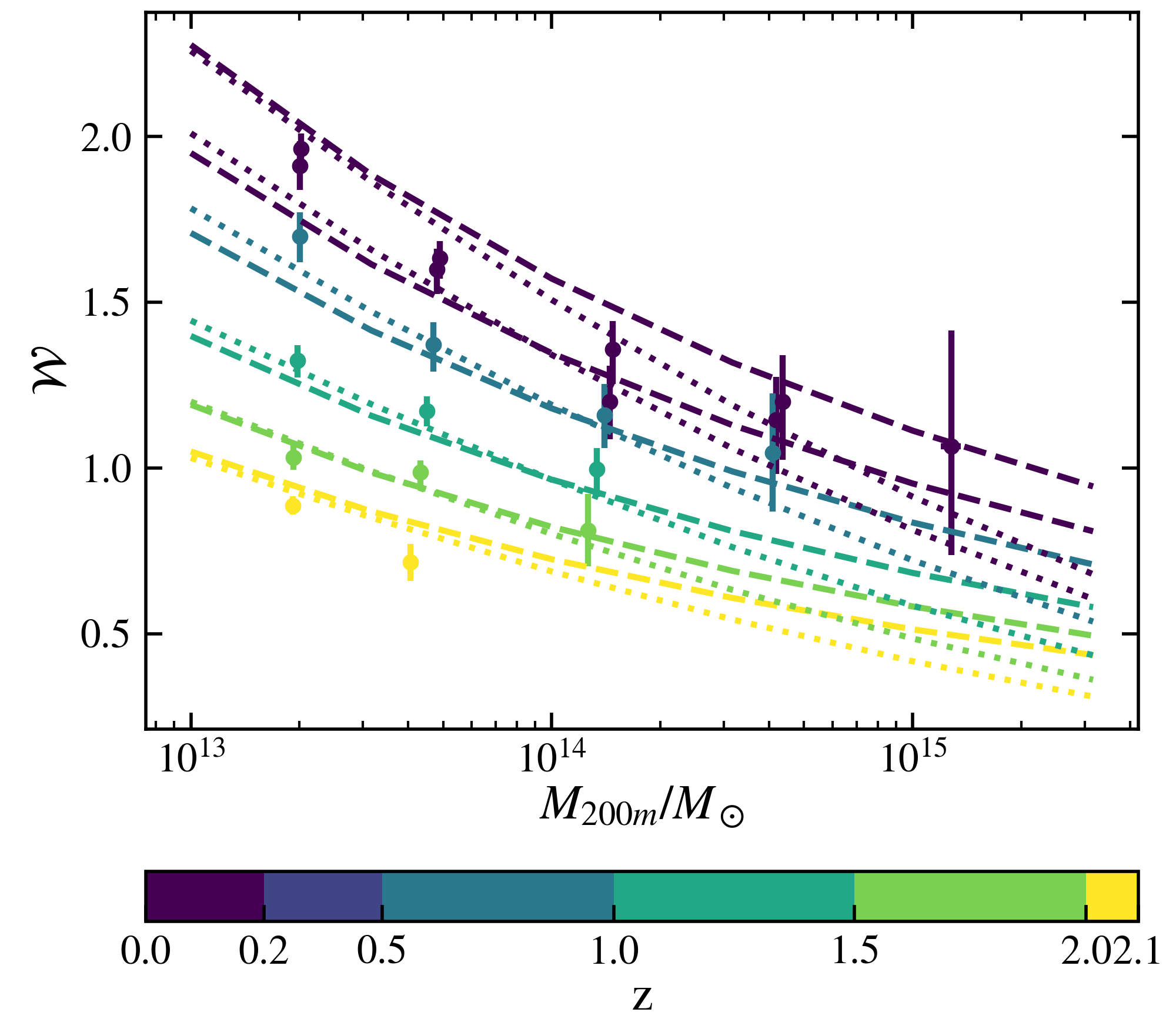}
    \caption{Comparison between the splashback width fitting functions. The dashed line represents the fitting in Equation \ref{eq:W_mass} obtained by fitting the mass data directly ($\chi_{\nu}^2=2.2$) while the dotted line represents the fitting in Equation \ref{eq:W_ph_M} derived from the relationship between the width and peak height ($\chi^2_{\nu}=1.2$). }
    \label{fig:W_compare}
\end{figure}
\newpage
\subsubsection{Peak height}
\label{sec:results_ph}
\input{results/pH}
\subsubsection{Formation time}
\label{sec:results_formz}
\input{results/formz}

%% file: results/conc.tex
\begin{figure}
    \centering
    \includegraphics[width=\linewidth]{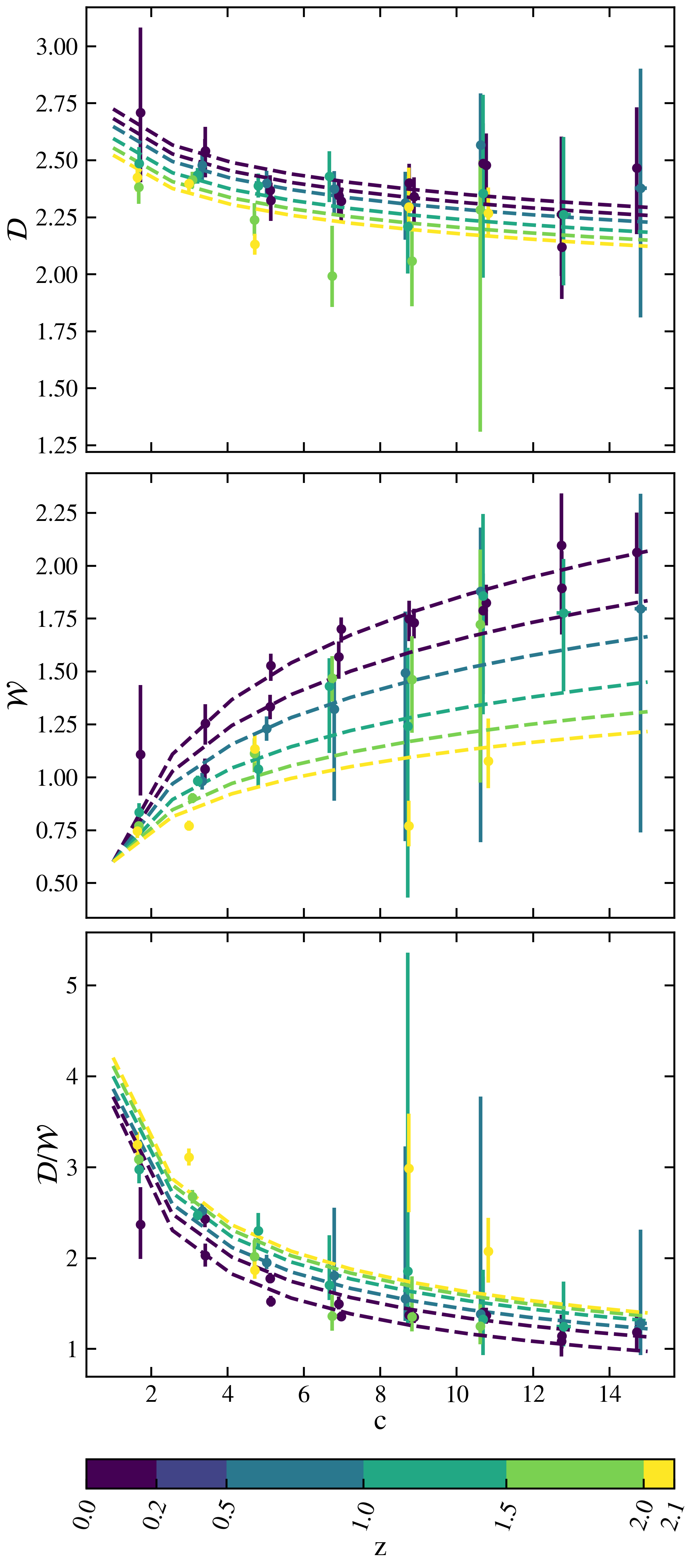}
    \caption{\textit{Top:} Depth of splashback features as a function of halo NFW profile concentration and binned by redshift shown by the colours. 
    \textit{Middle}: Width of splashback features as a function of halo NFW profile concentration with redshift shown by the colour. 
    \textit{Bottom}: Depth/Width of splashback features as a function of halo NFW profile concentration with redshift shown in colours. 
    }
    \label{fig:conc_depth}
\end{figure}
\begin{figure}
    \centering
    \includegraphics[width=\linewidth]{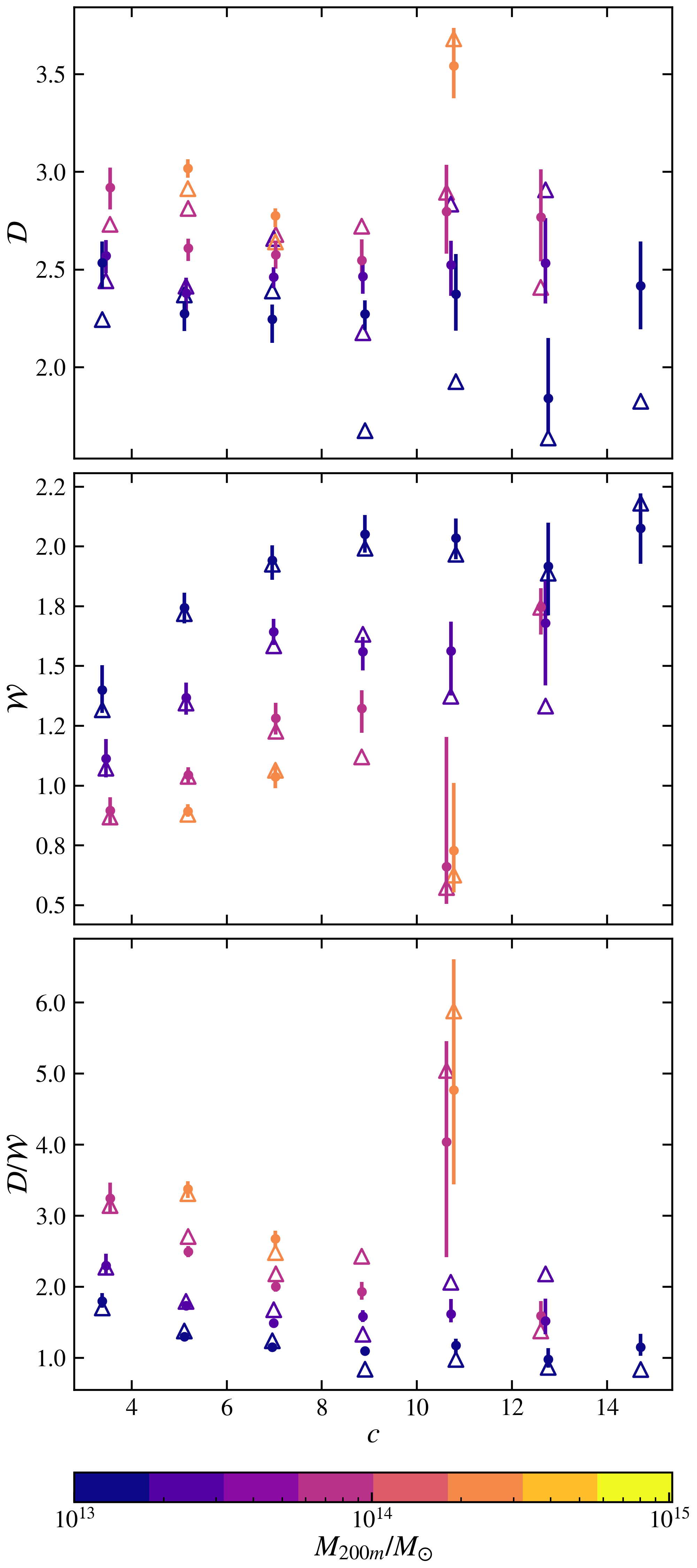}
    \caption{At $z=0$, haloes are binned by mass and concentration (colour-coded).
The dots with error bars show results from fitting the DK14 profile (Equation~\ref{eq:profile}), while the hollow triangles show results obtained by substituting the scale radius $r_s$ in the DK14 profile with that from the NFW fit (Equation~\ref{eq:NFW}).
    \textit{Top:} Depth of splashback features as a function of halo concentration. 
    \textit{Middle:} Width of splashback features as a function of halo concentration. 
    \textit{Bottom}: Depth/width of splashback features as a function of halo concentration.
    The splashback features are more significant with lower concentration and higher halo mass. 
    The splashback features are more pronounced for haloes with lower concentration and higher mass.
    The close agreement between points and triangles demonstrates that these trends are insensitive to the specific definition of concentration, confirming the robustness of our conclusions to the fitting method.
    }
    \label{fig:conc_mass_bin}
\end{figure}

The concentration of a halo depends on various properties of a halo, including mass, redshift, and formation history~\citep[e.g.][]{Bullock2001,Wechsler2002, L_pez_Cano_2022}.
For example, 
haloes that collapse and virialize earlier in cosmic time tend to be more concentrated, as the universe was denser at their formation time \citep{Wechsler2002}.
Consequently, low-mass haloes, which typically form earlier, have higher concentrations. Massive haloes, forming later, accrete matter more recently and remain less centrally concentrated.
This trend also depends on cosmology: for instance, an earlier matter domination epoch leads to earlier structure formation and higher concentrations at fixed mass \citep[e.g.][]{L_pez_Cano_2022,Shen2024-emd}. 
These dependencies may influence the properties of the splashback feature.

We bin haloes at each redshift by concentration. By fitting the NFW profile to individual haloes, we extract the parameters $\rho_s$ and $r_s$ given in Equation \ref{eq:NFW}. 
We calculate the concentration of individual haloes and select haloes with concentrations between $c=1$ and $c=19$ at intervals of $2$.
We then calculate the splashback features of the stacked set of haloes in each bin. 

In Figure \ref{fig:conc_depth}, we show the depth (top), the width (middle) and the depth-to-width ratio (bottom) as a function of concentration for various redshifts shown by the colours.
The depth decreases steadily with concentration while the width gradually increases with the concentration. 
The depth of haloes with various concentrations and redshifts is around $\mathcal{D}=1.8 \sim 3.0$. 
The width initially increases quickly from $\mathcal{W}=0.50$ to $\mathcal{W}=2.5$ until concentration $c\sim19$ and then decreases slightly from $c=2.5$ to $c=2.0$.
We fit the splashback depth and width with the functions (depth: $\chi^2_{\nu}=2.5$; width: $\chi^2_{\nu}=2.07$)
\begin{align}
    \mathcal{D}(c, z)&=Ac^B(z+1)^C\notag \\
    &=(3.03\pm0.07)c^{-0.16\pm0.01}(z+1)^{-0.13\pm0.01}\;,
    \label{eq:D_conc_z}
\end{align}
%
\begin{align}
    \mathcal{W}(c,z)&=A\left[\log c\right](z+1)^B+D\notag \\
    &=\left(0.54\pm0.03\right)\left[\log c\right](z+1)^{-0.79\pm0.18}+(0.60\pm0.06).
    \label{eq:W_conc_z}
\end{align}
The width equation can be effectively re-expressed as
\begin{equation}
    e^\mathcal{W}=e^Dc^{A(z+1)^B}.
\end{equation}
When taking the splashback depth-to-width ratio, in the bottom panel of Figure \ref{fig:conc_depth}, it consistently decreases with the concentration.
The depth-to-width ratio decreases slowly with overall variation $\Delta \mathcal{D}/\mathcal{W}=3.0$. 
By taking the ratio between Equations \ref{eq:D_conc_z} and \ref{eq:W_conc_z}, we apply the fittings to the splashback depth-to-width as a function of concentration and redshift
\begin{align}
    \mathcal{D}/\mathcal{W}(c,z)&=\frac{Ac^B(z+1)^C}{D\left[\log c\right](z+1)^E+F},
\end{align}
with $\chi_{\nu}^2=2.9$.

The fitting function reveals that haloes with the same concentration have larger depth and width at present, while the splashback depth-to-width ratio is smaller. 
This indicates that the splashback feature is more significant at higher redshift, given the same concentration. 
This is because, under fixed concentration, the inner density profile remains fixed. As a result, any variation in the splashback feature reflects changes in the outer halo dynamics, such as the halo mass accretion rate.
Haloes at earlier times accrete mass more rapidly \citep{van2002universal, zhao2003growth}, and we observe that haloes at higher redshift exhibit smaller splashback widths and thus more distinct splashback features, as shown in Figure~\ref{fig:accret_depth}.
Therefore, when halo profiles are varied primarily by their outer dynamics — as is the case when binning by concentration — splashback features become less significant at lower redshift.
As a halo increases in concentration, the depth decreases while the width increases. 
This matches the previous fittings.
As a halo increases in concentration, the splashback features get less distinctive. 
This is because high-concentration haloes are typically older and have experienced slower accretion in recent times. 
In \citet{reed2005evolution}, it was shown that haloes which form earlier develop steeper density profiles, as reflected by higher concentrations. This correlation arises because the inner structure of a halo is set during its early accretion phase and remains largely unchanged thereafter.
Meanwhile, haloes with lower accretion rates tend to have higher concentrations. 
As we have seen in Section \ref{sec:results_accret}, slowly accreted haloes at the present have less distinct splashback features. 
The particles in such haloes are more dynamically mixed and have more time to virialise, which smooths out the sharp density drop at the splashback radius. 
In contrast, low-concentration haloes have recently acquired rapid mass accretion. This leads to a sharper, more pronounced splashback feature due to less mixing and more radial orbits \citep{diemer2014dependence, adhikari2014splashback, more2015splashback}.

Furthermore, in order to verify that the splashback feature dependence on concentration is independent of the density profile fitting model chosen, we select haloes at present, bin them by halo mass $M_{200\text{m}}$ and concentration $c$ defined in Section \ref{sec:def_conc}, and fit them with different fitting models.
Within each bin, we fit the halos with both the NFW profile in Equation \ref{eq:NFW} and the DK14 profile in Equation \ref{eq:profile}, and then extract the parameter $r_s$ which relates to the concentration.
The $r_s$ in the NFW profile corresponds to the scale radius where the logarithmic slope of the density profile equals $-2$, whereas the $r_s$ in the DK14 profile is a more general shape parameter that sets the transition between the inner Einasto-like component and the steepening outer component and is coupled to additional parameters such as $\beta$ and $\gamma$.
By substituting the parameter $r_s$ measured from Equation \ref{eq:NFW} with the $r_s$ in Equation \ref{eq:profile}, we can simulate an additional DK14 profile. 
Therefore, by comparing the measured splashback features from the DK14 profile (dots) with the additional DK14 profile with $r_s$ from the NFW profile (hollow triangles) in Figure \ref{fig:conc_mass_bin}, we observe that the two types of splashback features match well with each other. 
The similarity between the points and the triangles in the figure reveals the insensitivity of the splashback features to the exact definition of concentration and the density profile fitting model used.
This demonstrates that the observed concentration dependence of the splashback features—predicted using the NFW profile—is consistent with measured splashback features based on the DK14 model.
The validation of the NFW-based fitting indicates that splashback features are strongly influenced by the inner halo profile, particularly through the scale radius.

%% file: results/pH.tex
The peak height of a halo provides a description of how large a density fluctuation the halo represents and is often used to relate halo mass to the underlying matter density field. 
It has a one-to-one correspondence with the mass of a halo and redshift.
Halos sharing the same $\nu$ correspond to equally rare initial density peaks, so they can be compared across redshift as systems at similar evolutionary stages and in similar environments.

Since we see redshift dependence in the relation between the depth and width of our splashback features with other properties such as mass, we now investigate whether this dependence persists with quantities like peak height that incorporate redshift in their definition.
\citet{diemer2014dependence} found that the logarithmic slope profiles of dark matter haloes vary significantly with peak height. 
At $1\leq\nu\leq1.5$, the logarithmic slope minimum reaches -3, while at $\nu\geq3.5$, the minimum reaches -4. For all peak heights, the slope minimum is more negative at lower redshift.
To quantify these changes, in Figure \ref{fig:feature_peakheight}, we show the depth and width of the splashback feature as functions of the peak height and redshift of the halo.

We calculate the peak height of the halo using the code \textsc{Colossus} \citep{Diemer2018} with the mass $M_{\rm200m}$, redshift of the snapshot and cosmology of MTNG.
We stack by peak height $\nu$ in 9 bins between $0<\nu<5$ and fit the resulting logarithmic slope of the density profile to identify the splashback feature.

There is an increase of the depth with the peak height.
At $z=0$, the depth increases from $\mathcal{D}=2.0$ to $\mathcal{D}=2.8$ while at $z=2$, the depth increases from $\mathcal{D}=2.2$ to $\mathcal{D}=2.7$ at $\nu<5$.
This is consistent with the findings in \citet{diemer2014dependence} since the depth is more significant with larger peak height at lower redshift.
However, there is significant scatter in the relationship between the depth of the splashback feature and the peak height, so the splashback depth has no strong dependence on peak height.
We apply a single power law to fit the splashback depth ($\chi_{\nu}^2=1.1$)
\begin{align}
    \mathcal{D}(\nu,z)&=A\nu^B + Cz^D \notag
    \\&= (1.89\pm0.02)\nu^{0.44\pm0.03}-(0.41\pm 0.04)z^{1.1\pm0.1}.
    \label{eq:D_ph_z}
\end{align}
The fitting form in Equation \ref{eq:D_ph_z} treats the dependencies on peak height and redshift as separable, i.e., no cross-terms are included. 
Thus the fitted parameters describe independent power-law scalings with $\nu$ and $z$, and no correlation between the two is assumed.

The width, on the other hand, has a very strong dependence on the peak height.
Above a peak height of $\nu\sim1$, the width steadily decreases with peak height, with a width of $\mathcal{W}\sim2.0$ near a peak height of $\nu\sim1$ and a width of $\mathcal{W}\sim1.0$ near a peak height of $\nu\sim4$.
The splashback width fitting function indicates that it is approximately linearly dependent on the peak height ($\chi_{\nu}^2=1.06$)
\begin{equation}\label{eq:W_ph_z}
    \mathcal{W}(\nu)=A\nu^B =(2.56\pm0.05)\nu^{-0.87\pm0.04}.
\end{equation}

Therefore, the peak height itself includes the information of redshift in the splashback features, which is considered to be a more invariant measure of the splashback features. 
This redshift dependence can be explicitly revealed by re-expressing the fitting function in terms of mass and redshift.
Since there is a one-to-one correspondence between halo mass and halo peak height, we use this equation to predict the splashback width dependence on halo mass.
Based on Equation \ref{eq:peak_height}, we have the splashback width in terms of mass
\begin{equation}\label{eq:W_ph_M}
    \mathcal{W}(M,z)=A\left[\frac{\delta_c}{\sigma(M, z)}\right]^B=(2.6\pm0.1)\left[\frac{\delta_c}{\sigma(M, z)}\right]^{-0.90\pm0.05}.
\end{equation}
By inputting the halo mass $M_{200\text{m}}$ and redshift into \textsc{Colossus}, we can compute the halo peak height. We compare the predictions with the data displayed in Figure \ref{fig:mass_depth_z}, and we get similar fitting in dotted lines with accuracy with $\chi^2_M=1.2$.
The result validates that Equation \ref{eq:W_ph_z} accurately captures the splashback width dependence on halo mass and redshift through peak height. 

There are a few explanations for this relation. 
The higher width of the splashback feature at low $\nu$ is possibly due to the scatter in environments at low peak height.
Some of these haloes may be located near massive neighbours in dense regions while others are isolated.
This causes a higher variance in the stacked profiles near $R_{\rm sp}$ and causes the width to increase.
\citet{diemer2014dependence} reported a much larger scatter for low-$\nu$ ($\nu<1$) haloes near massive neighbours compared to isolated ones, with $\frac{R_{\rm sp}}{R_{\rm200m}}$ spreading across approximately $0.3$.

There are two additional factors that may contribute to this trend.  One is that haloes with larger peak heights are in general more massive. As we saw in Section \ref{sec:mass}. Therefore, the positive correlation between the splashback depth and peak height and the negative dependence of splashback width on peak height are expected. Secondly, haloes with larger peak heights have lower concentrations; the transition of the profile from $r^{-3}$ to $r^{-1}$ happens at larger radii, which makes the splashback width smaller as well at higher peak heights. 

\begin{figure}
    \centering
    \includegraphics[width=\linewidth]{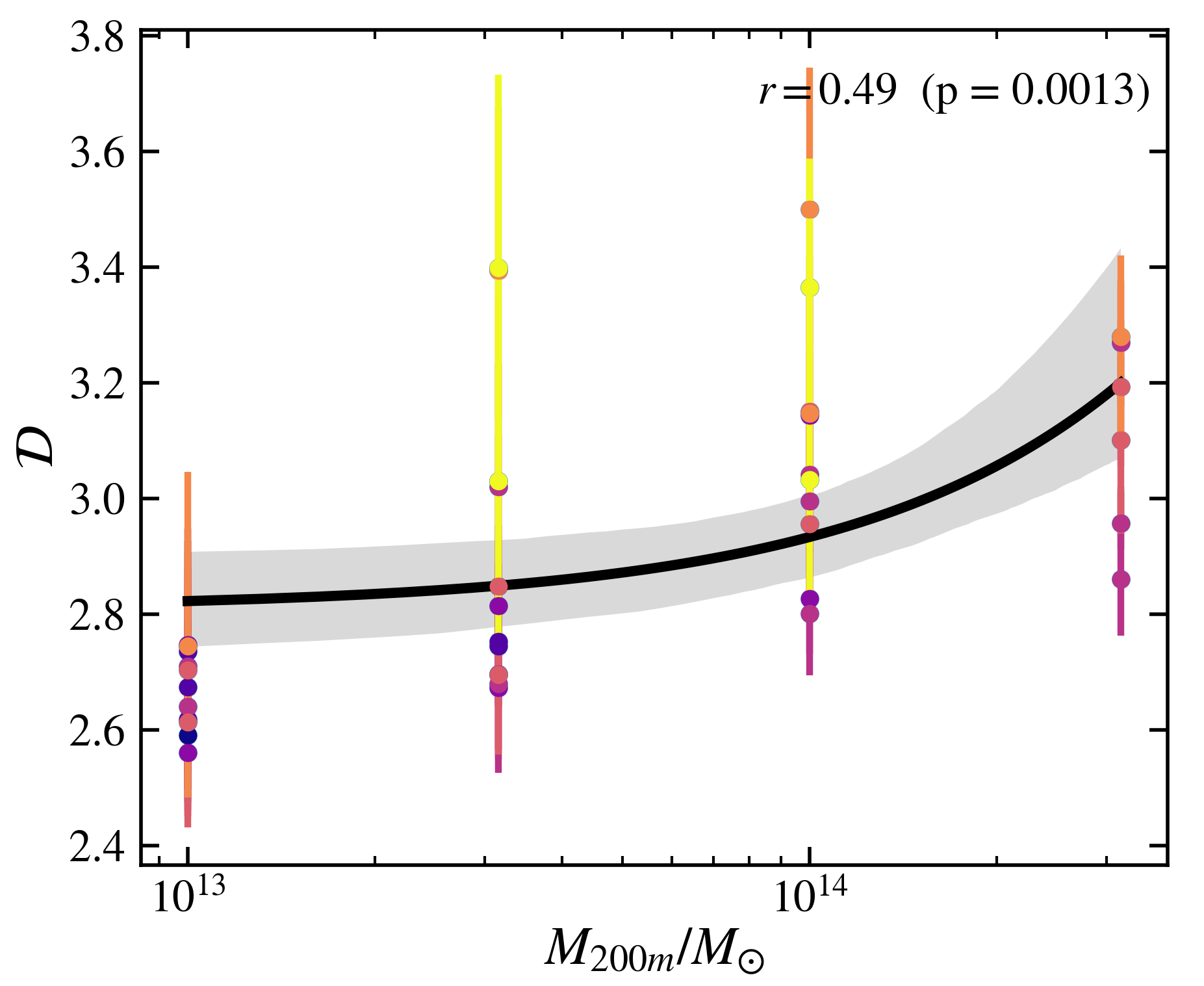}
    \includegraphics[width=\linewidth]{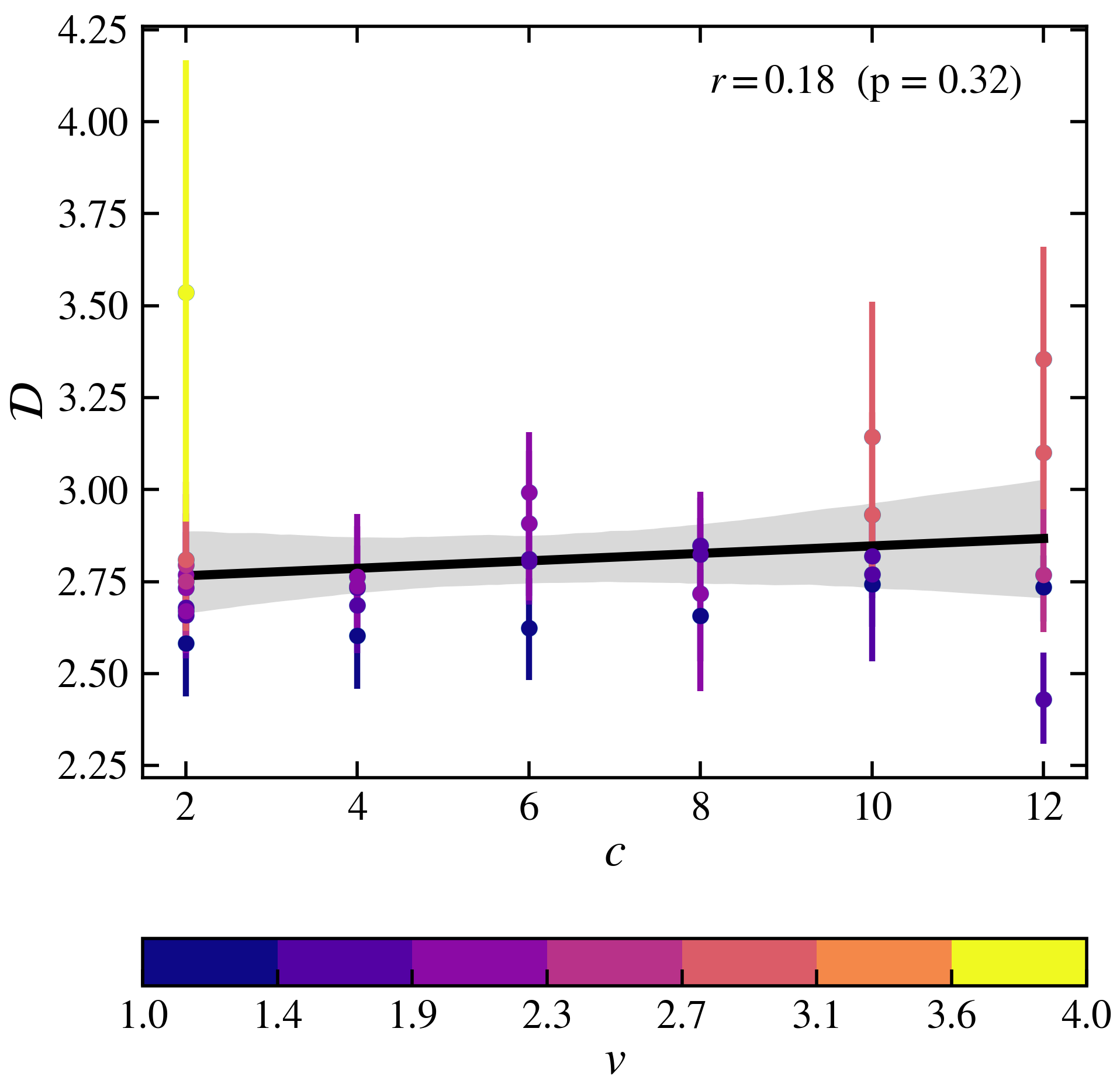}
    \caption{The splashback depth as a function of halo mass (\textit{Top}) and concentration (\textit{Bottom}) binned by peak height. 
    Regardless of binning by peak heights, the splashback depth demonstrates a stronger correlation with mass than the concentration.}
    \label{fig:pH_mass_conc_bins}
\end{figure}

In order to determine which factor, halo mass or concentration, has a more significant impact on the splashback features in terms of peak height, we bin haloes by both mass ($10^{13} \leq M_{200\text{m}} \leq 10^{15}\ M_\odot$, with $\Delta \log M = 0.5$) and concentration ($2 \leq c \leq 12$, with $\Delta c = 2$), and use \textsc{scipy} and \textsc{seaborn} to compute the correlations between splashback features and these two variables.
We use the depth as an example in Figure \ref{fig:pH_mass_conc_bins}. The plot shows that the splashback depth has a stronger correlation with mass than the concentration when binned by peak height. Therefore, the contribution from halo mass is more significant and has a greater effect on the splashback features as a function of peak height. 

Overall, when composing the splashback depth and width, the depth-to-width ratio demonstrates consistent dependence on the halo peak height with fitting functions ($\chi_{\nu}^2=0.72$)
\begin{align}
    \mathcal{D/W}(\nu,z)&= A\nu^B + Cz^D \notag
    \\&=(0.77\pm0.01)\nu^{1.21\pm0.03}-(0.27\pm0.04)z^{1.1\pm0.2}.
    \label{eq:DW_ph_z}
\end{align}
Equations \ref{eq:D_ph_z}, \ref{eq:W_ph_z} and \ref{eq:DW_ph_z} indicate that by binning the haloes by peak height, the splashback feature dependence on peak height and redshift can be separated.
The splashback features get more and more significant with increasing halo peak height and time, which is consistent with the findings of the splashback features' dependence on halo mass.

%% file: results/formz.tex
\begin{figure*}
    \centering
    \includegraphics[width=\linewidth]{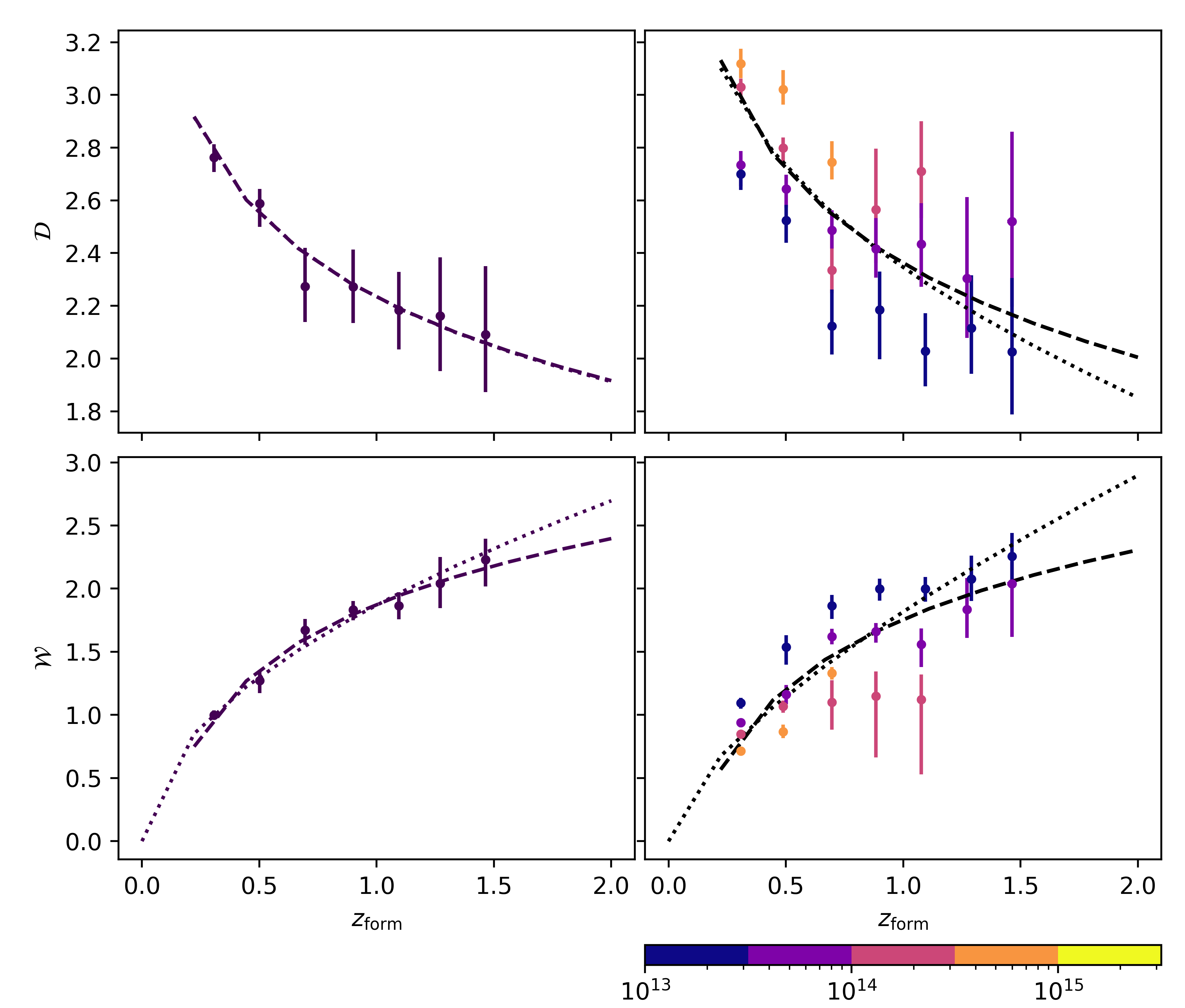}
    \caption{Haloes at $z=0$ are binned by the formation time in the left panel and binned by the formation time and the halo mass in the right panels. The formation time is defined as the time the halo reaches half of its $z=0$ mass. 
    \textit{Top:} Depth of splashback features as a function of halo formation time. 
    The dashed line represents the linear fitting while the dotted line represents the fitting derived from halo mass. 
    \textit{Bottom:} Width of splashback features as a function of halo formation time. 
    The dashed line represents the logarithmic fitting while the dotted line represents the fitting derived from halo mass. 
    }
    \label{fig:formz_mass_bin}
\end{figure*}
\begin{figure}
    \centering
    \includegraphics[width=\linewidth]{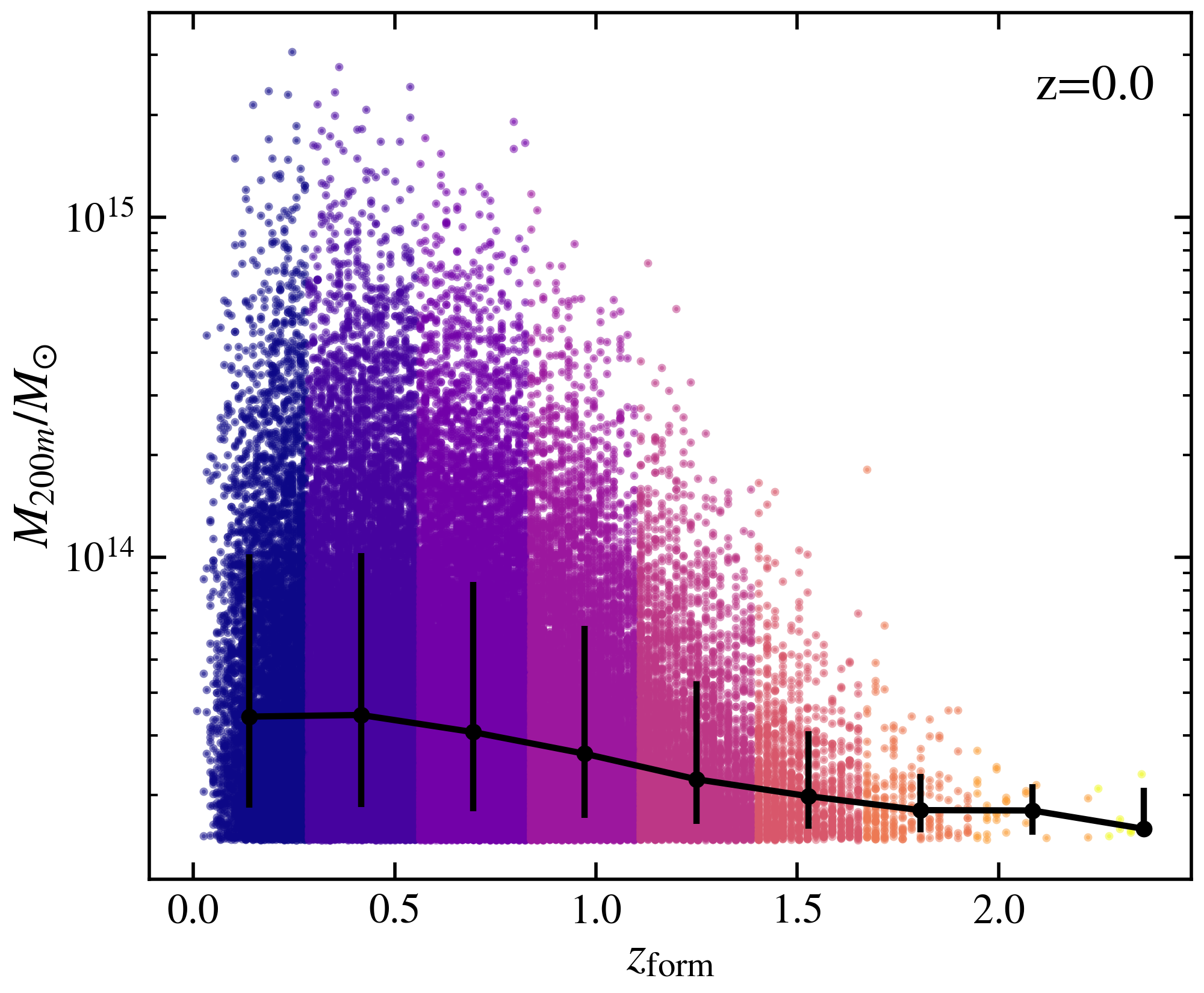}
    \caption{The mass-formation time relations at $z=0$. 
    The black line indicates the median mass in each formation time bin. 
    The colours correspond to the formation time bins used in the previous fitting for visualisation.
    Because more massive haloes tend to form later, the dependence of splashback features on formation time is expected to show trends opposite to those seen with halo mass.
    }
    \label{fig:formz_mass}
\end{figure}
Current-day properties are influenced by the formation history of the halo, so we now examine the dependence of the splashback feature on the halo formation time.
This effect — namely, the dependence of halo density profiles on formation time — has also been observed in \citet{Wechsler2002} and \citet{Diemer2025}.
This is also related to the dependence on concentration.
Similar dependence is also found in \citet{zhao2003mass, ludlow2014mass}, where early-forming haloes are shown to develop higher concentrations and steeper inner density profiles due to rapid early mass accretion.
These results provide indirect evidence that the dependence of splashback features on halo formation time arises from the connection between a halo’s assembly history and its internal structure.

In order to provide a quantitative measure of the splashback features on halo formation time, we define the formation time of a halo as the time when the halo reaches half of its current mass (at $z=0$). 
We select haloes at $z=0$ with formation time less than $z=1.5$ and bin the haloes with formation time by $\Delta z_{\text{form}}=0.2$.
Above z=1.5, most haloes have not yet reached half their $z=0$ mass. As seen in Figure \ref{fig:formz_mass}.
In the left panel of Figure \ref{fig:formz_mass_bin}, the splashback depth decreases and the splashback width increases with an increase in halo formation time.
The splashback depth decreases approximately linearly from 2.8 to 2.0 while the splashback width increases from 1.0 to 2.2 with a gradually slowed rate.
It is shown that for haloes formed at $z_{\text{form}}<0.75$, the splashback width decreases at a relatively faster rate comparing to the haloes formed at $z_{\text{form}}>0.75$.

We implemented the fitting functions of the depth and width of the splashback features to quantify the changes
\begin{align}
    \mathcal{D}(z_{\text{form}})&=a\log(z_{\text{form}})+b \notag\\
    &=(-1.05\pm0.06)\log(z_{\text{form}})+(2.23\pm0.02)
    \label{eq:D_formz}
\end{align}
with $\chi_{\nu}^2=0.23$, and
\begin{align}
    \mathcal{W}(z_{\text{form}})&=a\log(z_{\text{form}})+b\notag\\
    &=(0.75\pm0.04)\log(z_{\text{form}})+(1.88\pm0.03)
    \label{eq:W_formz}
\end{align}
with $\chi_{\nu}^2=0.45$.
Both the depth and the width vary with halo formation time logarithmically.
Therefore, the haloes formed earlier in time have less distinct halo boundaries and thus shallower and wider splashback features. 
Older haloes tend to have smaller masses as they grow more slowly than massive haloes \citep{McBride2009,zhao2009accurate}. 
We plot the mass distribution with halo formation time in Figure \ref{fig:formz_mass} to also verify this argument. 
It is shown that haloes with lower formation time tend to have higher median mass.
As the splashback depth increases with halo mass, it should decrease with halo formation time. 
This matches the observation in Figure \ref{fig:mass_depth_z}.

Furthermore, as the splashback feature in terms of formation time is consistent with the result from halo mass in Section \ref{sec:mass}, we implement the mass model from \citet{Wechsler2002} and modify it such that the growth rate of halo mass is proportional to the halo current mass. 
Therefore, the halo mass fitting equation in terms of redshift is
\begin{equation}
    M=M_0 e^{-\alpha M_0z}
\end{equation}
where $M_0$ is halo's present mass $M_{200\text{m}}$ and $\alpha$ is a fitting parameter. 
In this model, the formation time is calculated as
\begin{equation}\label{eq:zform_half}
    z_{\text{form}}=\frac{C}{M_0}, \,\,\, C=\frac{\log2}{\alpha}.
\end{equation}
By substituting this expression into Equations \ref{eq:D_mass} and \ref{eq:massW} at $z=0$ respectively, we get the fitting function of splashback depth in terms of formation time ($\chi_{\nu}^2=0.29$)
\begin{align}
    \mathcal{D}(z_{\text{form}})&=A\left[\log C-\log\left(z_{\text{form}}\right)\right]^B,\, C=\frac{\log2}{\alpha}\;,
\end{align}
and the splashback width ($\chi_{\nu}^2=0.97$)
\begin{align}
    \mathcal{W}(z_{\text{form}})&=A\left[\log C-\log\left( z_{\text{form}}\right)\right]^B.
\end{align}
The best-fit parameters show strong degeneracy, with large amplitude $A$ and steep negative slope $B$, implying the model is not robust and well constrained by the data.
The fit is shown as dotted lines in Figure \ref{fig:formz_mass_bin}.
It is shown that both fitting functions of the splashback depth and the splashback width have good matching with the simulation data.
The two fittings coincide between $0.1\leq z_{\text{form}}\leq1.0$, and the fitting with the mass model tends to have smaller depth and larger width at $z_{\text{form}}\geq1.0$. This means the mass model predicts more distinct splashback features at relatively large $z_{\text{form}}$. 
Two fittings verify the consistency between the splashback mass and formation time dependence, and they are most accurate between $0.1\leq z_{\text{form}}\leq1.0$. Within this region, the fittings are relatively reliable. 
Beyond that, which equations match better with either simulations or observations is yet to be known. 

Taken together, the analyses of both halo concentration and formation time consistently point to a common physical interpretation: haloes that form earlier, and thus possess higher concentrations, tend to exhibit less distinct splashback features. 
This is likely because early-forming haloes have already established dense inner profiles, leaving little contrast between the virialised and infalling regions to generate sharp splashback signatures. 
In contrast, haloes that experience more recent growth accumulate matter predominantly in their outer regions, which affects less the splashback width.
Therefore, the halo assembly history influences the splashback features by shaping the inner density profile of the halo.
This is distinct from the results presented in previous sections where haloes at higher redshifts have more distinct splashback features.
In this case, haloes at $z=0$ but that reached half their mass at earlier times have more distinct splashback features.

%% file: conclusions.tex
In this work, we have explored the dark matter haloes' splashback feature dependence on halo physical properties, such as mass, accretion rate and halo formation history, using the MillenniumTNG hydrodynamic simulations. 
We compare their mutual influence on the halo splashback features and how the splashback features dynamically evolve with time by setting these physical properties as control variables. 
We have found that the halo splashback features get more prominent with halo mass and redshift. 
Quantities that have a strong correlation with halo mass, such as the peak height and concentration, also demonstrate similar influence on the halo splashback depth and width.
Through the lens of halo assembly history, we find that splashback width is largely shaped by factors that influence the inner density profile, such as concentration and formation time.
Therefore, we provide the analytical fitting equations of the splashback features in terms of these quantities and redshift, aiming to provide a more accurate measure of the halo splashback features.
However, the splashback features have less dependence on the instantaneous properties, such as accretion rate and the time of halo relaxation.
The depth and width record the halo formation's history instead of capturing the dynamical features at halo boundaries. 
This complements information from the splashback radius itself.
These findings seem to have little dependence on the baryonic physics by comparing the simulation results between the dark matter-only and the hydrodynamic runs of MillenniumTNG with the same resolution. 
We summarise our results as follows:
\begin{itemize}
    \item In Figure \ref{fig:mass_depth_z}, we show the depth and width as a function of mass and redshift. The depth increases with mass while the width decreases. We provide a fitting function for these relations in Equations \ref{eq:D_mass} and \ref{eq:W_mass}.
    \item In Figure \ref{fig:accret_depth}, we show the depth and width as a function of accretion rate and redshift. The trend at higher redshifts is less clear than with mass, so we provide a fitting function only at redshift $z=0$ in Equations \ref{eq:D_acc_z0_pow} and \ref{eq:D_acc_z0_lin} for the depth and Equation \ref{eq:W_acc_z0} for the width.  We also fit these quantities as a function of redshift for haloes stacked by accretion rate in Equations \ref{eq:W_acc_z} and \ref{eq:W_acc_a}.
    \item We do not find a clear trend of the depth and width of the splashback feature with the timing of the most recent major merger.  We show this in Figure \ref{fig:mergerz_submass_bin}.
    \item We examine the depth, width and ratio of the two as a function of concentration in Figures \ref{fig:conc_depth} (also as a function of redshift) and \ref{fig:conc_mass_bin} (also binned by mass).  While the depth decreases weakly with concentration, the width increases and the ratio decreases with concentration.  The decrease of the ratio with concentration is significantly less pronounced at lower redshift. We provide fitting functions for the width and ratio as a function of concentration and redshift in Equations \ref{eq:D_conc_z} and \ref{eq:W_conc_z}.
    \item The depth, width and the ratio of the two vary with peak height $\nu$ and redshift as shown in Figure \ref{fig:feature_peakheight}.  The depth increases, the width decreases, and the ratio increases with $\nu$.  We provide fitting functions in Equations \ref{eq:D_ph_z}, \ref{eq:W_ph_z} and \ref{eq:DW_ph_z}.
    \item Finally, we show that the depth and width of the splashback feature depend on the formation time of the halo in Figure \ref{fig:formz_mass_bin}.  We explore this relation only at redshift $z=0$ and define formation time as the time a halo reached half its present-day mass.  The depth decreases with formation time while the width increases.  Some of this dependence is due to the relationship between present-day mass and formation time, but the trend remains even when splitting the sample by mass in addition to formation time.  We provide fitting functions in Equations \ref{eq:D_formz} and \ref{eq:W_formz}.
\end{itemize}

Importantly, we found that the relationship between the width and the peak height, $\mathcal{W}\propto\nu^{-0.87}$ as described in Section \ref{sec:results_ph}, to be particularly informative.
Using this relation largely removes any redshift dependence and predicts the mass dependence.
The depth also has a particularly strong dependence on the mass of the halo, following $\mathcal{D}\propto(\log_{10}M)^{2.8}$ as described in Section \ref{sec:results_mass}.

The splashback feature may also be able to probe aspects of a halo that we did not explore here, such as the nature of dark matter.
For example, \citet{Banerjee2020} showed that dark matter self-interactions cause the slope of the stacked density profile to become slightly shallower compared to CDM, but the position of $R_{\rm sp}$ does not exhibit a significant shift.
It may also be possible to further examine the features we did explore by binning stacks by multiple features like accretion rate and formation time to probe the effects of halo history and environment.
We present here just an initial study and hope to expand on the more promising features in the future.
Projection effects can also affect these features, although effects on the depth and width of the feature can be mitigated with proper fitting \citep{Sun2025}.
Observational quantities, like galaxy number density profiles or intracluster light, should also be examined to determine how well they trace these trends in the dark matter haloes.
In addition, many of the features, like the merger history or accretion rate, may be affected by stacking haloes rather than studying individual objects.
It may be possible to examine objects more individually by stacking by solid angle within a given halo.
These unexplored aspects highlight opportunities for future work.

%% file: appendix_fitting.tex
\section{Varying the density profile fitting function}
\label{apx:fitting}

\citet{Diemer2023} updated the fitting function proposed in \citet{diemer2014dependence} to consist of only infalling and orbiting terms:
\begin{align}
    \rho(r) &= \rho_{\rm orbit} + \rho_{\rm infall} \nonumber \\
    \rho_{\rm orbit} &= \rho_s \exp\left[-\frac{2}{\alpha}\left(\left(\frac{r}{r_s}\right)^\alpha-1\right) - \frac{1}{\beta}\left(\left(\frac{r}{r_t}\right)^\beta-\left(\frac{r_s}{r_t}\right)^\beta\right)\right] \nonumber \\
    \rho_{\rm infall} &= \rho_m \left[ \delta_1 \left[ \left(\frac{\delta_1}{\delta_{\rm max}}\right)^{\frac{1}{\xi}} + \left(\frac{r}{R_{200}}\right)^{\frac{s}{\xi}}\right]^{-\xi} + 1\right]\;.
    \label{eq:D23}
\end{align}

The logarithmic derivative of this function is then
\begin{align}
    \deriv{\log\rho(r)}{\log r} &= \frac{r}{\rho(r)} \deriv{\rho(r)}{r} = \frac{r}{\rho(r)}\left[\deriv{\rho_{\rm orbit}}{r} + \deriv{\rho_{\rm infall}}{r} \right] \notag\\
    \deriv{\rho_{\rm orbit}}{r} &= \rho_{\rm orbit}\left[-\frac{2}{r_s}\left(\frac{r}{r_s}\right)^{\alpha-1} - \frac{1}{r_t}\left(\frac{r}{r_t}\right)^{\beta-1}\right] \nonumber\\
    \deriv{\rho_{\rm infall}}{r} &= \frac{\rho_m}{R_{200}} \delta_1 \left[ \left(\frac{\delta_1}{\delta_{\rm max}}\right)^{\frac{1}{\xi}} + \left(\frac{r}{R_{200}}\right)^{\frac{s}{\xi}} \right]^{-\xi-1} \left(\frac{r}{R_{200}}\right)^{\frac{s}{\xi}-1}.
\end{align}
In this parametrisation, $\rho_s$ is the density at the scale radius $r_s$, and $r_t$ is the truncation radius.
The exponential parameters $\alpha$ and $\beta$ describe the inner slope and the sharpness of truncation respectively.
In the infalling term, $\delta_1$ is the density normalisation at $R_{200}$, $\delta_{\rm max}$ is the central overdensity, and $s$ is the power-law slope.

We fix the parameter $\xi$ at 0.5 as suggested in \citet{Diemer2023}, and $R_{200}$ is $1$ in our normalised profiles.
When fitting, we follow the suggested ranges in \citet{Diemer2023}:
$10<\rho_s/\rho_m<10^7$, $0.01<r_s/R_{200}<0.45$, $0.5<r_t/R{200}<3$, $0.03<\alpha<0.4$, $0.1<\beta<10$, $1<\delta_1<100$, $10<\delta_{\rm max}<2000$ and $0.01<s<4$.

We stack halos by mass as we did in Section \ref{sec:results_mass} and fit the splashback features as a function of mass and redshift. 
The fitting functions in Equations \ref{eq:D_mass} and \ref{eq:W_mass} also apply to the new halo profile in \citet{Diemer2023}.
The new fitted parameters are (depth: $\chi_{\nu}^2=1.4$; width: $\chi_{\nu}^2=1.8$)
\begin{align}
\mathcal{D}\left(M\right)=A\left(\log_{10}M\right)^B=(2.7\pm1.4)\times 10^{-3}\left(\log_{10}M\right)^{2.6\pm0.3};
\end{align}
\begin{align}
    \mathcal{W}\left(M, z\right)&=A\left(\log_{10}M\right)^B(z+1)^C\notag \\
    &=(8.0\pm1.2)\times10^5\left(\log_{10}M\right)^{-5.0\pm0.5}(z+1)^{-0.72\pm0.03} \; .
\end{align}
The comparison of fitting parameters is summarised in Table \ref{tab:compare}.
\begin{table}[]
    \centering
    \begin{tabular}{ccc}
        Parameters & \citet{diemer2014dependence} & \citet{Diemer2023} \\
        \hline
        $A_{\mathcal{D}}$ & $(1.8\pm0.7)\times 10^{-3}$ & $(2.7\pm1.4)\times 10^{-3}$\\
        $B_{\mathcal{D}}$ & $2.8\pm0.3$ & $2.6\pm0.3$\\
        $A_{\mathcal{W}}$ & $(8.5\pm1.9)\times 10^5$& $(8.0\pm1.2)\times 10^5$\\
        $B_{\mathcal{W}}$ & $-5.0\pm0.6$ & $-5.0\pm0.5$\\
        $C_{\mathcal{W}}$ & $-0.71\pm0.04$ & $-0.72\pm0.03$\\
    \end{tabular}
    \caption{
    Comparison of splashback feature fitting parameters derived from the \citet{diemer2014dependence} and \citet{Diemer2023} halo density profiles, illustrating the robustness of the fitted functional form.}
    \label{tab:compare}
\end{table}

It is shown that the variation in the fitted parameters remains within the error bars, and the splashback features preserve the fitted functional form.
This confirms that the fitted equations are robust and do not depend on the specific functional form of the density profiles but primarily on the shape of the profile.

The consistency of the splashback feature across different density profile models suggests that it is not sensitive to the specific form of the fitting function but instead represents an intrinsic quantity governed by the physical properties of dark matter haloes.
This robustness provides a foundation for the subsequent analyses in the main text, where splashback features are examined as functions of other halo properties beyond mass and redshift.

%% file: appendix_dmo.tex
\section{Comparison between the dark matter-only and hydrodynamic simulations}
\label{apx:hydro-DM}

To assess the impact of baryonic physics on the splashback features, we compare the results from the hydrodynamic and dark matter-only simulations of the MillenniumTNG suite.
This comparison tests whether the presence of baryons significantly alters the measured splashback features, and whether the fitted functional forms remain valid across different simulation types.

The splashback depth has strongest dependence on the halo mass. 
We follow the same procedures as described in Section \ref{sec:results_mass}, stacking halos by mass and fitting the splashback features as functions of mass and redshift using the dark matter-only simulation.
The same functional forms defined in Section \ref{sec:fitting}, namely Equations \ref{eq:D_mass} and \ref{eq:W_mass}, are used.
For the dark matter-only simulation, the fitted splashback depth and width are
(depth: $\chi_{\nu}^2=1.4$; width: $\chi_{\nu}^2=2.2$)
\begin{align}
\mathcal{D}\left(M\right)=A\left(\log_{10}M\right)^B=(1.8\pm0.8)\times 10^{-3}\left(\log_{10}M\right)^{2.8\pm0.3};
\end{align}
\begin{align}
    \mathcal{W}\left(M, z\right)&=A\left(\log_{10}M\right)^B(z+1)^C\notag \\
    &=(8.5\pm3.0)\times10^5\left(\log_{10}M\right)^{-5.0\pm0.6}(z+1)^{-0.71\pm0.06} \; .
\end{align}
The best-fit parameters from both simulation types are summarised in Table \ref{tab:AppenB}.
\begin{table}[]
    \centering
    \begin{tabular}{ccc}
        Parameters & Hydrodynamic & Dark matter-only \\
        \hline
        $A_{\mathcal{D}}$ & $(1.97\pm0.7)\times 10^{-3}$ & $(1.8\pm0.8)\times 10^{-3}$\\
        $B_{\mathcal{D}}$ & $2.8\pm0.3$ & $2.8\pm0.3$\\
        $A_{\mathcal{W}}$ & $(8.5\pm1.9)\times 10^5$& $(8.5\pm3.0)\times 10^5$\\
        $B_{\mathcal{W}}$ & $-5.0\pm0.6$ & $-5.0\pm0.6$\\
        $C_{\mathcal{W}}$ & $-0.71\pm0.04$ & $-0.71\pm0.06$
    \end{tabular}
    \caption{
    Comparison between the fitting parameters of splashback feature in terms of halo mass and redshift using both hydrodynamic and dark matter-only simulations, illustrating the robustness of the fitted functional form.}
    \label{tab:AppenB}
\end{table}

Similarly, the splashback width has strongest dependence on the peak height, so we bin the haloes by the peak height as we did in Section
\ref{sec:results_ph}.
We fit the splashback features with the same functional forms as Equations \ref{eq:D_ph_z}, \ref{eq:W_ph_M} and \ref{eq:DW_ph_z}. 
Using the dark matter-only simulations, they become (depth: $\chi_{\nu}^2=3.8$; width: $\chi_{\nu}^2=7.5$)
\begin{align}
    \mathcal{D}(\nu,z)&=A\nu^B+Cz^D\notag \\
    &=(1.97\pm0.03)\nu^{(0.41\pm0.03)}-(0.49\pm0.06)z^{(1.0\pm0.1)};\\
    \mathcal{W}(\nu)&=A\nu^B=(2.5\pm0.2)\nu^{-(0.77\pm0.07)}.
\end{align}
%
We observe that the splashback features as a function of peak height and redshift is more consistent in the hydrodynamic than the dark matter-only simulations.
The comparison between the fitting parameters of are shown in Table \ref{tab:AppenB_ph}.
\begin{table}[]
    \centering
    \begin{tabular}{ccc}
        Parameters & Hydrodynamic & Dark matter-only \\
        \hline
        $A_{\mathcal{D}}$ & $1.89\pm0.02$ & $1.97\pm0.03$\\
        $B_{\mathcal{D}}$ & $0.44\pm0.03$ & $0.41\pm0.03$\\
        $C_{\mathcal{D}}$ & $-0.41\pm0.04$ & $-0.49\pm0.06$\\
        $D_{\mathcal{D}}$ & $1.1\pm0.1$ & $1.0\pm0.1$\\
        $A_{\mathcal{W}}$ & $2.56\pm0.05$& $2.5\pm0.2$\\
        $B_{\mathcal{W}}$ & $-0.87\pm0.04$ & $-0.77\pm0.07$
    \end{tabular}
    \caption{
    Comparison between the fitting parameters of splashback feature in terms of peak height and redshift using both hydrodynamic and dark matter-only simulations, illustrating the robustness of the fitted functional form.}
    \label{tab:AppenB_ph}
\end{table}

The fitted parameters remain highly consistent within error bars, and the splashback features preserve the same mass and redshift dependencies across both simulation types.
This indicates that the presence of baryonic physics has negligible impact on the splashback features derived from spherically averaged halo profiles.
The result confirms that splashback features are primarily governed by the underlying gravitational dynamics of dark matter halos, and that the fitting procedure is robust to variations in baryonic modelling.